%
%
%
\def\@{{\char'100}}

\long\def\abstract#1{\bigskip{\advance\leftskip by 2true cm
\advance\rightskip by 2true cm\eightpoint\centerline{\bf
Abstract}\everymath{\scriptstyle}\vskip10pt\vbox{#1}}\bigskip}
\long\def\resume#1{{\advance\leftskip by 2true cm
\advance\rightskip by 2true cm\eightpoint\centerline{\bf
R\'esum\'e}\everymath{\scriptstyle}\vskip10pt \vbox{#1}}}

\def\references{\bigbreak\centerline{\sc
References}\medskip\nobreak\bgroup
\def\ref##1&{\leavevmode\hangindent 15pt
\hbox to 15pt{\hss\bf[##1]\ }\ignorespaces}
\parindent=0pt
\everypar={\ref}\par}
\def\endreferences{\egroup}
\long\def\authoraddr#1{\medskip{\baselineskip9pt\let\\=\cr
\halign{\line{\hfil{\Addressfont##}\hfil}\crcr#1\crcr}}}
\def\Subtitle#1{\medbreak\noindent{\Subtitlefont#1.} }
%
%
\newif\ifrunningheads
\runningheadstrue
\immediate\write16{- Page headers}
\headline={\ifrunningheads\ifnum\pageno=1\hfil\else\ifodd\pageno\rightheadline
\else\leftheadline\fi\fi\else\hfil\fi}
\def\rightheadline{\sc\hfil\RightHeadText\hfil}
\def\leftheadline{\sc\hfil\LeftHeadText\hfil}

\hyphenation{Harnad Neumann}
%
%
\immediate\write16{- Fonts "Small Caps" and "EulerFraktur"}  
%
%
%

\let\sc=\tensmc
%
%
\font\teneuf=eufm10  \font\seveneuf=eufm7 \font\fiveeuf=eufm5
\newfam\euffam \def\gr{\fam\euffam\teneuf}

\textfont\euffam=\teneuf \scriptfont\euffam=\seveneuf 
\scriptscriptfont\euffam=\fiveeuf
%
\edef\smatrix[#1\&#2\\#3\&#4]{\left({#1 \atop #3}\, {#2 \atop #4}\right)}

\def \smaller {\eightpoint}

\def \mt {\mapsto}
\def \ra {\rightarrow}

\def \lra {\longrightarrow}
\def \lmt {\longmapsto}
\def \a {\alpha}
\def \b {\beta}

\def \g {\gamma}
\def \G {\Gamma}
\def \k {\kappa}

\def \l {\lambda}

\def \r {\rho}
\def \s {\sigma}
\def \S {\Sigma}
\def \th {\vartheta}
\def \t {\tau}
\def \o {\omega}

\def \ss {\subset}

\def \mod{{\rm mod\,}}

\def\nchi{\hbox{\raise 2.5pt\hbox{$\chi$}}}
%
%

\def\grG{{\gr G}}

\def\nchi{\hbox{\raise 2.5pt\hbox{$\chi$}}}
%
%

\def\DD{{\cal D}}

%
%
		\def\bfA{{\bf A}}
		
		\def\bfC{{\bf C}}

		\def\bfI{{\bf I}}

		\def\bfP{{\bf P}}
		
		\def\bfR{{\bf R}}
		\def\bfS{{\bf S}}

		\def\bfZ{{\bf Z}}

%
%
\def\authorfont{\sc}
\font\eightrm=cmr8
\font\eightbf=cmbx8
\font\eightit=cmti8
\font\eightsl=cmsl8

\def\eightpoint{\let\rm=\eightrm \let\bf=\eightbf \let\it=\eightit
\let\sl=\eightsl \baselineskip = 9.5pt minus .75pt  \rm}

\font\titlefont=cmbx10 scaled\magstep2
\font\sectionfont=cmbx10
\font\Subtitlefont=cmbxsl10
\font\Addressfont=cmsl8
%
%
\def\Proclaim#1:#2\par{\smallbreak\noindent{\sc #1:\ }
{\sl #2}\par\smallbreak}
\def\Demo#1:#2\par{\smallbreak\noindent{\sl #1:\ }
{\rm #2}\par\smallbreak}
%
%
\immediate\write16{- Section headings}
\newcount\secount
\secount=0
\newcount\eqcount
\outer\def\section#1.#2\par{\global\eqcount=0\bigbreak
\ifcat#10
 \secount=#1\noindent{\sectionfont#1. #2}
\else
 \advance\secount by 1\noindent{\sectionfont\number\secount. #2}
\fi\par\nobreak\medskip} 
%
%
\immediate\write16{- Automatic numbering} 
\catcode`\@=11
\def\adv@nce{\global\advance\eqcount by 1}
\def\unadv@nce{\global\advance\eqcount by -1}
\def\nextnumber{\adv@nce}
%
%
\newif\iflines
\newif\ifm@resection
\def\onesec{\m@resectionfalse}
\def\moresec{\m@resectiontrue}
\moresec
\def\eq{\global\linesfalse\eq@}
\def\eqn{\global\linestrue&\eq@}
\def\nosubind@x{\global\subind@xfalse}
\def\newsubind@x{\ifsubind@x\unadv@nce\else\global\subind@xtrue\fi}
\newif\ifsubind@x
\def\eq@#1.#2.{\adv@nce
 \if\relax#2\relax
  \edef\loc@lnumber{\ifm@resection\number\secount.\fi
  \number\eqcount}
  \nosubind@x
 \else 
  \newsubind@x
  \edef\loc@lnumber{\ifm@resection\number\secount.\fi
  \number\eqcount#2}
 \fi
 \if\relax#1\relax
 \else 
  \expandafter\xdef\csname #1@\endcsname{{\rm(\loc@lnumber)}}
  \expandafter
  \gdef\csname #1\endcsname##1{\csname #1@\endcsname
  \ifcat##1a\relax\space
  \else
   \ifcat\noexpand##1\noexpand\relax\space
   \else
    \ifx##1$\space
    \else
     \if##1(\space
     \fi
    \fi
   \fi
  \fi##1}\relax
 \fi
 \eq@@{\loc@lnumber}}
\def\eq@@#1{\iflines \else \eqno\fi{\rm(#1)}}
\def\m@th{\mathsurround=0pt}
%
%
\def\display#1{\null\,\vcenter{\openup1\jot
\m@th
\ialign{\strut\hfil$\displaystyle{##}$\hfil\crcr#1\crcr}}
\,}
\newif\ifdt@p
\def\@lign{\tabskip=0pt\everycr={}}
\def\displ@y{\global\dt@ptrue \openup1 \jot \m@th
 \everycr{\noalign{\ifdt@p \global\dt@pfalse
  \vskip-\lineskiplimit \vskip\normallineskiplimit
  \else \penalty\interdisplaylinepenalty \fi}}}
%
%
\def\displayno#1{\displ@y \tabskip=\centering
 \halign to\displaywidth{\hfil$
\@lign\displaystyle{##}$\hfil\tabskip=\centering&
\hfil{$\@lign##$}\tabskip=0pt\crcr#1\crcr}}
%
%
\def\cite#1{{[#1]}}
\catcode`\@=\active
%
\hyphenation{Sebbar}
%
\magnification=\magstep1
\hsize= 6.75 true in
\vsize= 8.75 true in 
%
%
\def\RightHeadText{Modular solutions to generalized Halphen equations}
\def\LeftHeadText{J. Harnad and J. McKay}
%
%
\leftline{solv-int/9804006
\hfill CRM-2536 (1998) \break} \bigskip 
\bigskip \bigskip
\centerline{\titlefont Modular Solutions to Equations}
\centerline{\titlefont of Generalized Halphen Type}
\bigskip
\centerline{\authorfont J.~Harnad and J.~McKay}
\authoraddr
{Department of Mathematics and Statistics, Concordia University\\
7141 Sherbrooke W., Montr\'eal, Qu\'e., Canada H4B 1R6, {\rm \eightpoint
and} \\ 
Centre de recherches math\'ematiques, Universit\'e de Montr\'eal\\
C.~P.~6128, succ. centre ville, Montr\'eal, Qu\'e., Canada H3C 3J7\\
{\rm \eightpoint e-mail}: harnad\@crm.umontreal.ca \quad
mckay\@cs.concordia.ca} 
\bigskip

\abstract{Solutions to a class of differential systems that generalize
the Halphen system are determined in terms of automorphic functions whose 
groups are commensurable with the modular group. These functions all
uniformize  Riemann surfaces of genus zero and have $q$--series with integral
coefficients.  Rational maps relating these functions are derived, implying
subgroup relations between their automorphism groups, as well as
symmetrization maps relating the associated differential systems.} 
\bigskip \baselineskip 14 pt

\section 1. Introduction 

    Differential equations satisfied by modular functions have been
studied since the time of Jacobi \cite{J}. Such equations arise naturally in
connection with second order Fuchsian differential operators whose monodromy
groups coincide  with the automorphism group of the function.
Recall that a  meromorphic function $f(\tau)$ defined on an open, connected 
domain $\DD$ in the Riemann sphere is said to be automorphic with  respect to
a group $\grG_f$ of linear fractional transformations
$$
T:\t \ra {a\t  + b \over c \t + d} \equiv T(\t), 
\qquad \pmatrix{a & b \cr c & d } \in  GL(2, \bfC), \eq LFT..
$$
if the domain $\DD$ is $\grG_f$ invariant and
$$
f(T(\tau))= f(\tau), \qquad \forall \ T\in \grG_f.  \eq..
$$
In this paper, the term {\it modular} will be applied to functions with
Fuchsian automorphism groups of the first kind that are {\it commensurable}
with the modular group $PSL(2,\bfZ)$ (i.e., whose intersection with the latter
is of finite index in both).  

  Using the terminology of \cite{Fo}, a ``simple'' automorphic function is one
without essential singularities at ordinary points, whose fundamental region
has a finite number of sides and which has a definite (finite or infinite)
limiting value at any parabolic point (cusp). The following standard theorem
gives the connection between linear second order equations and  simple
automorphic functions (cf. \cite{Fo}, Sec.~44, Theorem 15)
\Proclaim  Theorem 1.1:  If $f(\tau)$ is a nonconstant, simple automorphic
function, then the (multi--valued) inverse function $\t = \t (f)$ can be
expressed as the quotient of two solutions of a second order linear equation
$$
{d^2 y \over df^2} + R(f) y = 0,  \eq Fuchstand..
$$
where $R$ is an algebraic function of $f$. If $f$ has a single first order
pole in the fundamental region, then $R$ is a rational functions of $f$.

In fact, the solutions of \Fuchstand may be expressed, at least locally,  as
$$
y = {(A + B\t (f)) \over (\t')^{1\over 2}}.  \eq..
$$
Conversely, given any second order linear equation 
$$
{d^2y \over d f^2} + P {d y \over d f} + Q y = 0  \eq Fuchsgen..
$$
with rational coefficients $P(f), Q(f)$, singular points $( a_1, \dots
a_n, \infty)$, a basis of solutions  $(y_1, y_2)$ and a base point $f_0$, the
image of the monodromy representation
$$
\eqalign{
M: \pi_1(\bfP -\{a_1, \dots a_n, \infty\}) &\ra GL(2,\bfC) \cr
M: \g & \mt M_\g =: \pmatrix {a & b \cr c & d},}  \eq..
$$
defined up to global conjugation by
$$
\g : (y_1, y_2)\vert_{f_0} = (y_1, y_2)\vert_{f_0} M_\g,  \eq..
$$
determines a subgroup $\grG \ss GL(2, \bfC)$ that acts on the
ratio
$$
\t := {y_1 \over y_2} \eq..
$$
by linear fractional transformations \LFT. By a simple substitution of the
type
$$
y \ra \prod_{i=1}^n (f - a_i)^{\mu_i} y, \eq..
$$
eq.~\Fuchsgen can be transformed into the form \Fuchstand without changing the
projective class of the monodromy group; i.e., without changing the
$\grG$--action \LFT on $\t$.  If $\grG$ is Fuchsian and commensurable with
$SL(2,\bfZ)$, the inverse function $f=f(\t)$  is a modular function in the
above sense.

   If $R(f)$ is the resulting rational function in \Fuchstand, then $f=f(\t)$
satisfies the Schwarzian differential equation
$$
\{f, \t\}+ 2 R(f) f'^2  = 0,  \eq Schwarzftau..
$$
where
$$
\{f, \t\} := {f'''\over f'} -{3\over 2} \left({f''\over f'}\right)^2, 
\qquad  (f':={df\over d\t}) \eq..
$$
denotes the Schwarzian derivative \cite{H1, GS}.

  Perhaps the oldest example of such an equation involves the
square  of the elliptic modulus of the associated Jacobi elliptic functions
$\l(\t) = k^2(\t)$, which satisfies the Schwarzian equation \cite{H1}
$$
\{\l, \t\} + {\l^2 - \l + 1\over 2\l ^2(1 -\l)^2 }\l'^2 = 0. 
\eq lambdaSchwarz..
$$
The automorphism group in this case is the level $2$ principal congruence
subgroup  
$$
\G(2)  :=\left\{g =\pmatrix {a & b \cr c & d}\in SL(2,\bfZ),
\ g \equiv \bfI \ (\mod \ 2)\right\}.   \eq..
$$
The associated Fuchsian equation is the hypergeometric equation
$$
\l(1-\l) {d^2 y \over d \l^2} + (1-2\l) {dy\over d\l} - {1\over 4} y = 0, 
\eq lambdahypergeom..
$$
for which a basis of solutions is given by the elliptic ${1\over 2}$--period
integrals
$$
\eqalignno{
\k &= \int_0^1 {dt \over \sqrt{(1-t^2)(1- k^2 t^2)}} 
= {\pi\over 2} F({\scriptstyle{1\over 2},  {1\over 2}}; 1; k^2) 
\eqn ellipticinta.a. \cr  i\k' &= \int_1^{1\over k} {dt 
\over \sqrt{(1-t^2)(1- k^2 t^2)}}  = {i\pi \over 2}  F({\scriptstyle{1\over 2},
{1\over 2}}; 1;1- k^2) \eqn ellipticintb.b.}
$$
with
$$
\t = {i \k' \over \k}.  \eq..
$$

  An equivalent way of representing the Schwarzian equation \lambdaSchwarz,
due to Brioschi \cite{B}, is to introduce the functions
$$
w_1 :=  {1\over 2}{d\over d\t}\ln {\l'\over \l},\quad
w_2 :=  {1\over 2}{d\over d\t}\ln {\l'\over (\l-1)},\quad
w_3 :=  {1\over 2}{d\over d\t}\ln {\l'\over \l(\l-1)}.  \eq wHalphen..
$$
These satisfy the system
$$
\eqalign{
w_1' &=  w_1(w_2 + w_3) - w_2 w_3   \cr
w_2' &=  w_2(w_1 + w_3) - w_1 w_3    \cr
w_3' &=  w_3(w_1 + w_2) - w_1 w_2,  }  \eq Halphen..
$$
introduced by Darboux \cite{Da} in his study of orthogonal coordinate systems 
and solved by Halphen
\cite{Ha} with the help of hypergeometric functions. This system is referred
to in \cite{T, O1, O2} as the Halphen equations. The general solution
\cite{Ha} is obtained by applying an arbitrary $SL(2,\bfC)$ transformation
\LFT to the independent variable,  while transforming $(2w_1, 2w_2, 2w_3)$ as
affine connections 
$$
w_i \lra {1\over (c \tau +d)^2} w_i\circ T - {c \over c\tau + d}.  \eq
AffConnect..
$$
The system \Halphen has appeared in a number of recent contexts, including:
solutions of reduced self--dual Einstein \cite{GP} and Yang--Mills
\cite{CAC, T} equations, the $2$--monopole dynamical equations \cite{AH}, and
the WDVV equations of topological field theory \cite{Du}.

  A symmetrized version may be obtained  by considering the symmetric 
invariants formed from $(w_1, w_2, w_3)$, such as
$$
W:= 2(w_1 + w_2 + w_3),   \eq WChazy..
$$
which satisfies the Chazy equation \cite{C1,C2} 
$$
W'''=2 W W'' - 3 W'^2 .  \eq Chazy..
$$
The corresponding modular function is obtained by noting that, although the
action of the full modular group $\Gamma := PSL(2, \bfZ)$ upon $\l$ does not
leave it invariant, the quotient $\G / \G(2)$ by the normal subgroup $\G(2)$
is just the symmetric group $\bfS_3$, acting on $\lambda$ as the ``group of
anharmonic ratios'' \cite{Hi2}
$$
\l \mt \l,\ {1\over \l}, \ 1-\l, \ {1 \over 1-\l}, 
\ {\l\over\l-1}, \ {\l-1\over \l}.  \eq anharmgroup..
$$
These transformations correspond, respectively, to the following
modular transformations
$$
\t \mt \ \t, \ {\tau\over1 + \tau}, \ -{1\over \t}, \ -{1 \over 1+\t}, 
\ \tau + 1, \ {\tau -1\over \tau}.  \eq GtoGtwo..
$$
Symmetrization amounts to forming the ring of invariants, which
has a single generator that may be taken as Klein's $J$-function
$$
J := {4(\l^2 -\l +1)^3\over 27 \l^2 (\l-1)^2},  \eq Jlambda..
$$
whose automorphism group is $\G$. The corresponding Fuchsian equation is the
hypergeometric equation 
$$
J(1-J) {d^2y \over dJ^2} + \left({2\over 3} - {7\over 6} J\right) {dy\over dJ}
-{1\over 144} y = 0.   \eq..
$$
The associated Schwarzian equation,
$$
\{J, \t\}+ { 36J^2 - 41J + 32\over 72 J^2(J-1)^2} J'^2 = 0,  \eq SchwarzJ..
$$
was first studied by Dedekind \cite{De} in relation to the modular
properties of $J$ and $\lambda$. The resulting solution of the Chazy equation
\Chazy is  given \cite{C1, T} by

$$
W = {1\over 2}{d\over d\t} \ln{J'^6\over J^4(J-1)^3} 
={d\over d\t} \ln{\l'^3\over \l^2(\l-1)^2} 
 = {1\over 2}{d\over d\t}\ln \Delta,  \eq dlogDelta..  
$$
where $\Delta$ is the modular discriminant.

  If we introduce analogous Halphen variables for the $J$-function
$$
W_1 :=  {1\over 2}{d\over d\t}\ln {J'\over J},\quad
W_2 :=  {1\over 2}{d\over d\t}\ln {J'\over (J-1)},\quad
W_3 :=  {1\over 2}{d\over d\t}\ln {J'\over J(J-1)},  \eq H..
$$
these satisfy the system
$$
\eqalign{
W_1' = & {1\over 4} W_1^2 + {1\over 9} W_2^2  +{23\over 36}W_1 W_2
+{31\over 36}W_1 W_3
-{31\over 36}W_2 W_3 \cr
W_2' = & {1\over 4} W_1^2 + {1\over 9} W_2^2  +{23\over 36}W_1 W_2
-{41\over 36}W_1 W_3
+{41\over 36}W_2 W_3  \cr
W_3' = & {1\over 4} W_1^2 + {1\over 9} W_2^2  -{49\over 36}W_1 W_2
+{31\over 36}W_1 W_3
+{41\over 36}W_2 W_3.   }  \eq HalphenJeq..
$$
From the transformations \anharmgroup, it follows that the symmetrizing group 
$\bfS_3$ acts upon the Halphen variables $(w_1,w_2,w_3)$ by
permutations. Differentiating the identity \Jlambda, we deduce that the
quantities $(W_1, W_2, W_3)$ are related to $(w_1, w_2,w_3)$ as follows
$$
\eqalignno{
3W_1+2W_2+W_3 &= 2\s_1 =  W  \eqn JlambdaWa.a.\cr
W_1-W_3 &= -{4(\s_1^2 - 3 \s_2)^2\over 2\s_1^3-9 \s_1 \s_2 + 27\s_3} 
\eqn JlambdaWb.b. \cr
W_2-W_3 &= -{2\s_1^3-9 \s_1 \s_2 + 27 \s_3\over \s_1^2 - 3 \s_2}, 
\eqn JlambdaWc.c.}
$$
where
$$
\s_1:= w_1 + w_2 +w_3, \quad \s_2 := w_1 w_2 + w_2 w_3 + w_3 w_1,
\quad \s_3 := w_1 w_2 w_3
\eq..
$$
are the elementary symmetric invariants. In terms of these, the system
\HalphenJeq reduces to 
$$
\eqalign{
\s'_1 &= \s_2 \cr
\s'_2 &= 6\s_3 \cr
\s'_3 &=  4\s_1\s_3  - \s_2^2}  
\eq  Halphensymm..
$$
(cf. \cite{O1}), which is equivalent to the Chazy equation \Chazy.

   The modular solutions to these systems may also be
represented as logarithmic derivatives of the null theta functions
$\vartheta_2(\tau)$, $\vartheta_3(\tau)$, $\vartheta_4(\tau)$, (cf.~Appendix),
in terms of which the elliptic modular function $\l(\tau)$ has
the well known representation \cite{WW}
$$
\l(\t) = {\vartheta_2^4 \over \vartheta_3^4} = 1 - {\vartheta_4^4 
\over \vartheta_3^4}.   \eq lambdatheta..
$$
From the differential identities satisfied by $\vartheta_1$, $\vartheta_2$,
$\vartheta_3$ (cf.~Appendix, eqs.~(A.19)--(A.21) and \cite{O1}), it follows 
that
$$
{\l' \over 1-\l} = i \pi \vartheta_2^4, \quad {\l'\over \l} 
= i \pi \vartheta_4^4,
\quad {\l'\over \l (1-\l)} = i\pi \vartheta_3^4,  \eq lamdatheta..
$$
so the solution of the Halphen system may be expressed as
$$
w_1= 2 {d \over d\tau} \ln\th_4, \qquad w_2= 2 {d \over d \tau} \ln\th_2, 
\qquad w_3= 2 {d \over d\tau} \ln\th_3. \qquad 
\eq..
$$ 

  Using \Jlambda, the $J$-function may  also be expressed rationally in terms
of the theta functions
$$
J = { (\vartheta_2^8 + \vartheta_3^8 + \vartheta_4^8)^3\over 
54\vartheta_2^8 \vartheta_3^8 \vartheta_4^8}.   \eq..
$$
Taking derivatives and applying the same differential identities, the $J$
Halphen variables $(W_1, W_2, W_3)$  may be expressed as
logarithmic derivatives of rational expressions in the $\vartheta$'s, and the
symmetrizing relations \JlambdaWa-\JlambdaWc interpreted as differential
relations satisfied by the theta functions. 

  The above examples have been generalized by Ohyama \cite{O2} to other
classes of second order equations of type \Fuchstand, both Fuchsian and
non--Fuchsian. For the Fuchsian case, the rational functions $R(f)$ may be
expressed as
$$
R(f) ={N(f)\over (D(f))^2},  \eq NoverD..
$$
where 
$$
D(f) = \prod_{i=1}^n (f-a_i)  \eq Denom..
$$
and $N(f)$ is a polynomial of degree $\leq 2n - 2$. 
The generalization of the Halphen variables is given by
$$
X_0 := {1\over 2} {d\over d \tau}\ln f', \quad 
X_i := {1\over 2}{d \over d \tau} \ln {f'\over (f- a_i)^2}, \qquad i=1, 
\dots n.  
\eq Ohyamavars..
$$
For $n>2$, the $X_i$'s again satisfy a system of first order equations
analogous to \Halphen and \HalphenJeq, with suitably defined quadratic forms
on the right, but they are also subject to a set of $n-2$ quadratic
constraints fixing the anharmonic ratios between any distinct set of four of
them. The underlying phase space is therefore still $3_\bfC$--dimensional and
may, for generic initial conditions, be identified with the $SL(2, \bfC)$
group manifold. Besides the above two cases, other explicit solutions were
given in terms of modular functions by Ohyama \cite{O3} for a system in which
the corresponding Fuchsian operator has four regular singular points and  the
automorphism group is $\G(3)$.

  For the case of Fuchsian operators with three regular singular points the
associated differential systems were already studied by Halphen
\cite{Ha}. For the general hypergeometric equation
$$
f(1-f) {d^2y \over df^2} + (c - (a + b + 1) f) {d y \over df} - a b y = 0, 
\eq Hypergeom..
$$
the corresponding rational function is
$$
R(f) = {1\over 4}\left({1-\l^2\over f^2} + {1-\mu^2\over (f-1)^2} + 
{\l^2 +\mu^2 -\nu^2 -1 \over f(f-1)}\right),  \eq Rhypergeom..
$$
where
$$
\l := 1-c, \quad \mu := c-a-b, \quad \nu := b-a  \eq..
$$
are the exponents at the regular singular points $(0,1,\infty)$,
which determine the angles \break $(\l \pi, \mu \pi, \nu \pi)$ at the
vertices. Introducing the variables
$$
W_1 := {1\over 2}{d\over d\t}\ln {f'\over f},\quad
W_2 := {1\over 2}{d\over d\t} \ln {f'\over (f-1)},\quad
W_3 := {1\over 2} {d\over d\t}\ln {f'\over f(f-1)},  \eq..
$$
and viewing them as functions of the ratio
$$
\t = {y_1\over y_2}
$$
of two independent solutions of \Hypergeom, these satisfy the general Halphen 
system
$$
\eqalign{
W_1' =& \mu^2 W_1^2 + \l^2 W_2^2 + \nu^2 W_3^2 \cr
&+(\nu^2 - \l^2 -\mu^2 +1)W_1 W_2
+(\l^2 - \mu^2 -\nu^2 +1)W_1 W_3
+(\mu^2 - \l^2 -\nu^2 -1)W_2 W_3 \cr
W_2' =& \mu^2 W_1^2 + \l^2 W_2^2 + \nu^2 W_3^2 \cr
&+(\nu^2 - \l^2 -\mu^2 +1)W_1 W_2
+(\l^2 - \mu^2 -\nu^2 -1)W_1 W_3
+(\mu^2 - \l^2 -\nu^2 +1)W_2 W_3   \cr
W_3' =& \mu^2 W_1^2 + \l^2 W_2^2 + \nu^2 W_3^2  \cr
&+(\nu^2 - \l^2 -\mu^2 -1)W_1 W_2
+(\l^2 - \mu^2 -\nu^2 +1)W_1 W_3
+(\mu^2 - \l^2 -\nu^2 +1)W_2 W_3 . }  \eq Halphengeneral..
$$

  Although this gives a construction, in principle, of the general solution to
such systems, the functional inversion involved is not in general globally 
well defined. Only if the parameters $(\l , \mu , \nu )$  are of the form 
$({1\over m}, {1\over n}, {1\over p})$, where $m, n$ and $p$ are integers or
$\infty$, is there a tessellation of the upper half plane by (triangular)
fundamental regions, and even then we cannot in general say much about the
explicit form of the inverse function. However, for a small number of special
cases which are described below, these again turn out to be modular functions
that can be given  explicit rational expressions in terms of null
$\vartheta$--functions or the Dedekind
$\eta$-function.

 Remarkably, the $J$-function also appears in connection with
``Monstrous Moonshine'' \cite{CN}, in that the $q = e^{2\pi i
\tau}$--coefficients of the Fourier expansion of $j :=12^3 J -744$ are the
dimensions of representations of the Monster simple sporadic group; i.e., the
traces on the identity element.  Like $j$, the  $q$--series with coefficients
given by traces on the other conjugacy classes turn out to also be
{\it Hauptmoduls}; i.e., each is the generator of a field of  meromorphic
functions of genus $0$. An alternative characterization of the elliptic 
modular function $j$ is the fact that the principal part of its $q$--series is
$q^{-1}$, together with its behaviour under the action of the Hecke operator;
namely,
$$
  T_n(j) = {1\over n}\sum_{0\leq b<d \atop ad=n }  
j\left({a\tau+b\over d}\right)
= P_{n,j}(j), \qquad \forall n \ge 1,  \eq..
$$
where $P_{n,f}(f)$ is the Faber polynomial \cite{Fa} of degree $n$.
The functions appearing as such character generators are included in a
larger class of Hauptmoduls, the {\it replicable functions} \cite{CN, FMN},
which are constructed using a generalization of the Hecke operator. Their
automorphism  groups all contain  a finite index subgroup of the type
$$
\G_0(N) :=\left\{\pmatrix {a & b \cr c & d } \in \ss SL(2,\bfZ), \ c\equiv 0 \
\mod N\right\}  \eq..
$$
(and hence their automorphism groups are all commensurable with
$\Gamma$.)  The maximal such $N$ is referred to as the level of the function. 
Like all Hauptmoduls, they also satisfy Schwarzian equations of the type
\Schwarzftau with rational $R(f)$ of the form
\NoverD, and hence each has an associated Fuchsian operator of the form 
\Fuchstand whose projectivized monodromy group is the automorphism group of
the function. 

  In the present work, we study the differential equations satisfied by
such Hauptmoduls and their corresponding generalized Halphen systems. In
particular, we consider the Hauptmoduls known as {\it triangular functions},
for which the tessellation of the upper half-plane is associated with a
triangular domain and a corresponding hyperbolic group generated by
reflections in its sides. The associated Fuchsian equations are
therefore of hypergeometric type. Since the automorphism groups appearing
are conjugate to subgroups of $PGL(2,Q)$, it follows that only
crystallographic angles
$(0,\pi,\pi/2,\pi/3,\pi/4,\pi/6)$ occur at the vertices. For the
$j$--function, the triangle has a cusp at $i\infty$ and angles of  $\pi/3$ at
$e^{i\pi/3}=J^{-1}(0)$ and $\pi/2$  at $i=J^{-1}(1)$, so  this is
denoted $({1\over 3}, {1\over 2}, 0)$. Since the functions considered have
$q$--expansions with  principal part ${q^{-1}}$ at $q=0$, there must
be a cusp at $\tau =i\infty$, and hence the angle $\nu$ always vanishes. There
are nine such triangular {\it arithmetic groups} \cite{Ta} with angles given
by: $(0,0,0)$, $(1/2,0,0)$, $(1/3,0,0)$, $(1/3,1/2,0)$, $(1/4,1/2,0)$,
$(1/6,1/2,0)$, $(1/3,1/3,0)$, $(1/4,1/4,0)$, $(1/6,1/6,0)$, and these are 
precisely the ones appearing in the list \cite{CN} of replicable functions 
with integral  $q$--series coefficients. 

   In the following sections, several examples will be given of differential
systems of generalized Halphen type whose solutions are completely determined 
in terms of replicable functions.  These will include all the arithmetic
triangular groups, and some further cases for which there are four singular
points in the associated Fuchsian equation (as well as one in which there are
$26$). Each such function  has a fundamental domain bounded by circular arcs
centered on the real axis, with a cusp at $i\infty$ and one of the
crystallographic angles $(0, \pi/2,\pi/3,\pi/4, \pi/6)$ at each remaining
vertex. By construction, each has a normalized $q$--series of the form
\cite{FMN}
$$
F(q) = {1\over q} + \sum_{n=1}^\infty a_n q^n,  \qquad q:= e^{2i\pi \t},
\eq qseriesnorm..
$$
with integer coefficients $a_n$. In the case of triangular functions, in order
to relate these to solutions of the hypergeometric equations, another standard
normalization is chosen by applying an affine transformation 
$$
F= \a f + \b,  \eq Affine..
$$ 
with constants $\a$ and $\b$ chosen to assign the values $(0,1,\infty)$ to
$f(\tau)$ at the vertices of the fundamental triangle. Since all these
functions are known explicitly \cite{FMN} in terms of the Dedekind eta
function $\eta(\tau)$ or the null theta functions $\vartheta_a(\tau)$, we are
able to give explicit solutions of the corresponding differential systems in
terms of logarithmic derivatives of the
$\eta$- or $\vartheta$-functions.

   Certain of these functions are expressible in terms of others
through a rational map, possibly composed with a transformation of the type
$\tau \mt \tau/n$ (where $n$ has prime factors $2$ or $3$). This implies that
there is a subgroup relation between their respective  automorphism groups, and
also that the differential system associated to the larger automorphism group
is a symmetrization of the one associated to the subgroup. An example of this
was seen in the  pair of elliptic modular functions $(J, \lambda)$, where the
rational expression \Jlambda implies the subgroup relation $\G(2) \ss \G$ and
the relations \JlambdaWa--\JlambdaWc between the correponding Halphen
variables.  In general, if $f$ satisfies the Schwarzian equation
\Schwarzftau and $g$ satisfies
$$
\{ g, \tau\} + 2 \tilde{R}(g) g'^2 = 0, \eq Schwarzgtau..
$$
then if $f$ can be expressed as a function of $g$
$$
f = Q(g), \eq..
$$
this will satisfy the Schwarzian equation
$$
\{Q, g\} + 2 R(Q(g)) Q'^2 = 2\tilde{R}(g).  \eq Schwarzfg..
$$ 

   Equivalently, if $Q(g)$ satisfies \Schwarzfg and $y(f)$ is a solution to
\Fuchstand, the composite function $\tilde{y}(g)=(Q')^{-{1\over 2}}y(Q(g))$ 
will satisfy
$$
{d^2 \tilde{y} \over dg^2} + \tilde{R}(g) \tilde{y} = 0.  \eq Fuchstandg..
$$
By choosing both $R(f)$ and $\tilde{R}(g)$ of the hypergeometric form 
\Rhypergeom, Goursat \cite{Go} found  rational transformations of
degree $\le 4$ relating various classes of hypergeometric functions. The
relation \Jlambda may be viewed as a composition of two such transformations,
of degrees $2$ and $3$, connecting the hypergeometric equations of types
$(a,b,c)= ({1\over 12}, {1\over 12}, {2\over 3})$ and   
$({1\over 2}, {1\over 2}, 1)$. In section 2b, transformations of Goursat's type
are used to relate the associated arithmetic triangular functions, and in
section 3b, similar transformations of degree $\le 4$ are given relating these
to certain $4$--vertex cases. Such relations are determined by finding
appropriately normalized rational solutions $f=Q(g)$ of the Schwarzian equation
\Schwarzfg or, equivalently, by equating the $q$--series of both sides up to a
finite number of terms.

 In all the cases treated, we give a tabular summary of the properties of the 
modular functions $f$,  including explicit rational expressions for them in 
terms of  the $\eta$- and $\vartheta$-functions. The general solutions to the 
Halphen systems \Halphengeneral and their multivariable generalizations are 
thus determined as logarithmic derivatives of $\eta$- or $\vartheta$-functions.
For  the hypergeometric cases (section 2a), we list the exponents $(\l,\mu,
\nu)$ determining the Halphen type system
\Halphengeneral, the corrresponding hypergeometric parameters $(a,b,c)$ and
the group elements fixing the vertices. For the four vertex cases (section
3a),  we give the locations
$(a_1, a_2, a_3)$ of the finite poles of the rational function $R(f)$; i.e., 
the values of $f$ at the vertices, the  group elements stabilizing the 
vertices and a representative of an associated  linear class of quadratic
forms in the dynamical variables that serves to uniquely define the
corresponding constrained $4$--variable system, as well as the rational
function $R(f)$ entering eqs.~\Fuchstand  and \Schwarzftau. Section 4 contains
a  discussion of the $n+1$ pole case and includes an example with  $26$
regular singular points, which is the largest number that appears, and
automorphism group of level $72$. In the general $n+1$ pole case, it is shown
that Ohyama's quadratically constrained
$n+1$ variable dynamical system \cite{O2} is equivalent to a
$3$--variable system defined on the $SL(2,\bfC)$ group manifold. 

   For both the triangular cases (section 2b),  and certain four vertex
cases (section 3b), we examine the various rational maps of degree $\le 4$
relating the different Hauptmoduls, together with the corresponding relations
between the associated automorphism groups. These maps define symmetrizations
for the function fields and the corresponding differential systems. Just as in
the $J$--$\lambda$ case \JlambdaWa--\JlambdaWc, the generalized Halphen
variables for the more symmetrical systems consist of rational invariants
formed from the less symmetrical one. Not all cases correspond to normal
subgroups, however, so the fibre of the quotient map is not necessarily a
Galois group, but rather the quotient of two Galois groups, corresponding to a
field extension whose automorphism group is the largest subgroup normal in
both the groups of the pair of Hauptmoduls.

\section 2.  Triangular replicable functions and symmetrization maps  

\Subtitle {2a. Solutions of Halphen type systems}
\smallskip
\nobreak

 In Table 1, we list all cases where the underlying Fuchsian equation is
of hypergeometric type \Hypergeom. The Halphen system is always of the form
\Halphengeneral, and therefore we list only the values of the constants
$(a,b,c)$  and $(\l, \mu, \nu)$ characterizing these systems. The notation
used to designate the corresponding automorphism groups and function fields is
that of \cite{FMN}. The triangular cases appearing are denoted: 1A, 2A, 3A,
2B, 3B, 4C, 2a, 4a and 6a, where the integer $N$ denotes the level. (The upper
case letters denote functions that are character generators for the Monster.)
There are further cases of replicable functions having triangular automorphism
groups, but these are all equivalent to one of the above under an affine
transformation that just relocates the two finite poles of the Fuchsian
equation \Fuchstand, together with a M\"obius transformation \LFT in $\tau$,
and therefore they are not listed separately. The $\eta$-function formulae
listed for the functions $f(\tau)$ normalized to take values $(0,1,\infty)$ at
the vertices are obtained from those given in \cite{FMN} for the functions $F$
normalized as in \qseriesnorm by applying the appropriate affine
transformation \Affine. The $\vartheta$-function formulae are deduced either
from standard identities relating the $\eta$- and
$\vartheta$-functions (cf.~Appendix), or by applying the rational maps
relating the different cases listed in sections 2(b) and 3(b). The
automorphism group, being the projective image of the monodromy group of the
associated hypergeometric equation, has three matrix generators
$\r_0,\r_1,\r_\infty$ stabilizing the vertices satisfying the relation
$$
\r_\infty \r_1  \r_0  = m \bfI,  \qquad m\in \bfR  \eq rhoprod..
$$
for some $m\neq 0$. Representatives of the projective class of these
generators may always be chosen to have integer entries, though not
necessarily unit determinant, so the constant $m$ in \rhoprod is an
integer.  In all cases the generator $\r_\infty$ stabilizing the cusp at
$i\infty$ is
$$
\r_\infty = \pmatrix{1 & 1 \cr 0 & 1}.  \eq rhoinf..
$$
and the two remaining elements $\rho_0$, $\rho_1$ are given in the table.
\bigskip
\medskip
\centerline{\bf{Table 1.  Triangular Replicable Functions}}\nobreak
\nobreak
\bigskip
\centerline{
\vbox{\tabskip=0pt \offinterlineskip
\def\tablerule{\noalign{\hrule}}
\halign to455pt{\strut#& \vrule#\tabskip=.5em plus2em&
 \hfil#\hfil & \vrule # &\hfil #\hfil & \vrule # &
 \hfil#\hfil & \vrule# & \hfil#\hfil & \vrule# &
 \hfil#\hfil & \vrule# & \hfil#\hfil & \vrule#
\tabskip=0pt\cr\tablerule
&& Name && $(a,b,c)$ && $(\lambda, \mu, \nu)$ && $\rho_0 \qquad \quad \rho_1$ 
&& $ F $ && $f(\tau)$   &\cr \tablerule
&& $\matrix{1A \cr \sim \G}$ && $({1\over 12},{1\over 12},{2\over 3})$ 
&& $({1\over 3},{1\over 2},0)$  &&
$\smatrix[ 0 \&-1 \\ 1 \& -1] $ \ $\smatrix[ 0 \&-1 \\ 1 \& \phantom{-}0] $
&& $ 1728 f -744$  && $J = {(\vartheta_2^8 + \vartheta_3^8 + 
\vartheta_4^8)^3\over  54 \vartheta_2^8 \vartheta_3^8 \vartheta_4^8}$
&\cr \tablerule
&& $2A$ && $({1\over 8}, {1\over 8}, {3\over 4})$ 
&& $({1\over 4}, {1\over 2}, 0)$ &&
$\smatrix[ 0 \&-1 \\ 2 \& -2] $ \ $\smatrix[ 0 \&-1 \\ 2 \& \phantom{-}0] $
 && $256 f -104$ &&  
${ \left(\vartheta_3^4 +\vartheta_4^4\right)^4 \over
16  \vartheta_2^8 \vartheta_3^4 \vartheta_4^4}$
&\cr\tablerule 
&& $3A$ && $({1\over 6}, {1\over 6}, {5\over 6})$ && 
$({1\over 6}, {1\over 2}, 0)$  && 
$\matrix{\cr\smatrix[ 0 \&-1 \\ 3 \& -3] $ \ $\smatrix[ 0 \&-1 \\ 3 \&
\phantom{-}0]\cr\phantom{m} }$
 && $108 f -42 $  &&  ${\left(\eta^{12}(\tau) + 27 \eta^{12}(3\tau)\right)^2
\over 108 \eta^{12}(\tau)\eta^{12}(3\tau)}$ 
&\cr\tablerule
&& $\matrix{2B \cr \sim\G_0(2)}$ && $({1\over 4}, {1\over 4}, {1\over 2})$ && 
$({1\over 2}, 0, 0)$  && 
$\matrix{\cr\cr\smatrix[ 1 \&-1 \\ 2 \& -1] $ \ $\smatrix[ -1 \& \phantom{-}0 
\\ \phantom{-}2 \& -1]\cr\cr\phantom{m}}$
 && $64 f -40  $  
&&  $\matrix{1 + {1\over 64}\big({\eta(\tau)\over
\eta(2\tau)}\big)^{\scriptscriptstyle 24}\cr = {\big(\vartheta_3^4(\tau) +
\vartheta_4^4(\tau)\big)^2\over
\vartheta_2^8(\tau)}} $
&\cr\tablerule
&& $\matrix{3B \cr \sim \G_0(3)}$ && $({1\over 3}, {1\over 3}, {2\over 3})$ && 
$({1\over 3}, 0, 0)$  && 
$\smatrix[ 1 \&-1 \\ 3 \& -2] $ \ $\smatrix[ -1 \& \phantom{-}0 \\ 
\phantom{-}3 \& -1]$
 && $ 27 f -15  $  &&  $1 + {1\over 27}\big({\eta(\tau)\over
\eta(3\tau)}\big)^{\scriptscriptstyle 12}$ 
&\cr\tablerule
&& $\matrix{4C^*\cr \sim \G_0(4)}$ && $({1\over 2}, {1\over 2}, 1)$ && 
$(0, 0, 0)$  && 
$\matrix{\cr\smatrix[ 1 \&-1 \\ 4 \& -3] $ \ $\smatrix[ -1 \& \phantom{-}0 \\ 
\phantom{-}4 \& -1 ]\cr\cr}$
 && $16 f - 8 $  &&  $\matrix{{1\over \l(2\tau)}
 = {\vartheta_3^4(2\tau) \over \vartheta_2^4(2\tau)}\cr 
=1 +{1\over 16}\big({\eta(\tau)\over\eta(4\tau)}\big)^{\scriptscriptstyle 8}
\cr }$ &\cr\tablerule
&& $2a$ && $({1\over 6}, {1\over 6},  {2\over 3})$ && 
$({1\over 3}, {1\over 3}, 0)$  && 
$\matrix{\cr\smatrix[ 2 \&-3 \\ 4 \& -4] $ \ $\smatrix[ 0 \&-1 \\ 4 \&
-2]\cr\phantom{m}}$
 && $24\sqrt{3}i(2 f - 1) $  &&  
${\sqrt{3}i\left( e^{\pi i/3}\th_3^4(2\tau) -\th_2^4(2\tau)\right)^3\over
9\th_2^4(2\tau)\th_3^4(2\tau)\th_4^4(2\tau)}$
&\cr\tablerule
&& $4a$ && $({1\over 4}, {1\over 4}, {3\over 4})$ && 
$({1\over 4}, {1\over 4}, 0)$  && 
$\matrix{\cr\smatrix[ 4 \&-5\\ 8 \& -8] $ \ $\smatrix[ 0 \&-1 \\ 8 \&
-4]\cr\phantom{m}}$
 && $-16i(2 f - 1)$  && $-{i\left(\th_3^2(2\tau) +
i \th_4^2(2\tau)\right)^4\over 8\th_2^4(2\tau)\th_3^2(2\tau)\th_4^2(2\tau)}$
&\cr\tablerule
&& $6a$ && $({1\over 3}, {1\over 3}, {5\over 6})$ && 
$({1\over 6}, {1\over 6}, 0)$  && 
$\matrix{\cr\smatrix[ 6 \&-7 \\ 12 \& -12] $ \ $\smatrix[ 0 \&-1 \\ 12 \&
-6]\cr\phantom{m}}$
 && $ 6\sqrt{3}i( 2f - 1)$  &&  
$-{\sqrt{3}i \left(\eta^6(2\tau) + 3\sqrt{3}i\eta^6(6\tau)\right)^2\over
36\eta^6(2\tau)\eta^6(6\tau)}$
&\cr\tablerule 
\hfil\cr}}}
\nobreak \vbox{
\noindent{\smaller  ${}^*${\it Remark:} Case ${\scriptstyle 4C}$ has the same
fundamental domain as the elliptic modular function ${\scriptstyle\l(\tau)}$, 
but with respect to the variable ${\scriptstyle \tau/2}$.  The automorphism
group is ${\scriptstyle \G_0(4)}$, which is conjugate to ${\scriptstyle \G(2)}$
under the map ${\scriptstyle T\lmt \smatrix[ 1 \& 0\\ 0 \& {1\over 2}]  T
\smatrix[1 \& 0 \\ 0 \& 2]}$.}} 
\vfill 
\bigskip

\Subtitle {2b. Rational maps between triangular cases}

\nobreak

   In the following, we catalogue the rational maps, analogous to \Jlambda,
relating the replicable functions listed above. Each map may be seen as
following from an identity relating different hypergeometric functions through
a rational change in the independent variable. Such identities were derived
systematically by Goursat \cite{Go}, and are labelled here according to the
numbering scheme of \cite{Go}. For each case, we have a pair $(f,g)$ of
function field generators, with $f$ a rational function of $g$,  normalized
consistently with the hypergeometric equations, so that the vertices of the
fundamental triangle are mapped to $(0, 1, \infty)$. In order to preserve the
standard ordering with angles $(\lambda, \mu, \nu)$ at the vertices with values
$(0, 1, \infty)$, we make use of the fact that $F(a,b,c;z)$ and
$F(a,b,1-c+a+b; 1-z)$ satisfy the same equation and compose, in some cases,
with the maps $f \mt 1-f$ or $g \mt 1-g$.  The corresponding pair of
functions, normalized as in \qseriesnorm, is denoted $(F,G)$ and the relevant
map is denoted  $G\mt F$. In order to preserve the form \qseriesnorm of the
$q$--series, it is necessary in some cases to compose the rational map with 
a map $\t \mt {\t\over 2}$ or  $\t \mt {\t\over 3}$ on the independent
variable. In these cases, the composed map is denoted $G'\mt F$. By taking
derivatives of the corresponding rational map relating $f$ to $g$, we also
obtain relations between the Halphen variables, such as that given in the case
of the pair
$(J,\l)$ by \JlambdaWa--\JlambdaWc. 

   To fix notation, the Halphen variables for the two cases will be denoted
$$
W_1 := {1\over 2}{d\over d\t}\ln {f'\over f},\quad
W_2 := {1\over 2}{d\over d\t} \ln {f'\over (f-1)},\quad
W_3 := {1\over 2} {d\over d\t}\ln {f'\over f(f-1)}  \eq WHalphenf..
$$
for the first element of the pair $(f,g)$, and
$$
w_1 := {1\over 2}{d\over d\t}\ln {g'\over g},\quad
w_2 := {1\over 2}{d\over d\t} \ln {g'\over (g-1)},\quad
w_3 := {1\over 2} {d\over d\t}\ln {g'\over g(g-1)}  \eq wHalpheng..
$$
for the second. Since $f$ is in the function field generated by $g$, it is
$\grG_g$--invariant; i.e., $\grG_g\ss\grG_f$. When $\grG_g$  is a normal
subgroup, $\grG_f$ acts on $g$ by rational tranformations, and on the Halphen
variables $(w_1, w_2, w_3)$ linearly.  Since this defines a finite group action,
the ring of invariants in the  $(w_1,w_2,w_3)$ variables is generated by $3$
elements, and the $(W_1,W_2,W_3)$ are uniquely determined as rational 
expressions of these. In all the cases listed below we give the
hypergeometric identity underlying the rational maps between the modular
functions and the maps, expressed in terms of both the $(f,g)$ normalizations
and the $(F,G)$ ones. When $\grG_g \ss \grG_f$ is a normal subgroup, we give
the symmetrizing group  $\bfS_f^g := \grG_f / \grG_g$ (i.e., the Galois group
of the field extension), the action of $\grG_f$ upon $g$  (analogous to the
$\bfS_3$ action \anharmgroup on the elliptic modular function $\l(\t)$), the
elementary symmetric invariants under this action, and the linear relations
determining $(W_1,W_2,W_3)$ in terms of the latter following from
differentiation of the rational map relating $f$ to $g$. When $\grG_g \ss
\grG_f$ is not normal, we list a symmetrization quotient $\bfS_f^h/\bfS_g^h$,
where $\bfS_f^h$ and $\bfS_g^h$ are the symmetrizing groups of the smallest 
field extension, with generator $h$, that is Galois for both. The symmetrizing
groups all turn out to be either a symmetric group $\bfS_n$, an alternating
group
$\bfA_n$ or a cyclic group $\bfZ_n$.

     We note that in each case, there exists at least one relation
involving a linear symmetric invariant of the form
$$
P W_1 + Q W_2 + R W_3 = p w_1 + q w_2 + r w_3  \eq linearsymminv..
$$
where $(P,Q,R,p,q,r)$ are all integers, and 
$$
P+Q+R=p+q+r =: k.  \eq..
$$
This is deduced from a relation of the form
$$
{f'^{P+Q+R}\over f^{P+R} (f-1)^{Q+R}} = M {g'^{p+q+r}\over g^{p+r}
(g-1)^{q+r}},  \eq eliminatesingular..
$$
where $(P,Q,R,p,q,r)$ are integers, which may always be found by simply 
choosing $(P,Q,R)$  so that the left hand side of \eliminatesingular has no
singularities at the vertices. The  resulting quantity is an analytic form of
weight $2k$, and its logarithmic derivative, given by \linearsymminv,
satisfies a third order equation analogous to the Chazy equation \Chazy.
\smallskip
\noindent $ 2a' \mt 1A $: 

\nobreak
\noindent Hypergeometric identity (Goursat (138)): 
$$
F\left({1\over 12}, {1\over 12}; {2\over 3}; 4x(1-x)\right)
= F\left({1\over 6}, {1\over 6}; {2\over 3}; x\right)  \eq..
$$
Rational map: 
$$
\eqalign{
F &= 984 + G^2 \circ (\t \mt \t/2), 
\cr
f &= 4g(1-g) \circ (\t \mt \t/2)}
\eq.. 
$$
Symmetrizing group: \quad $\bfS_{1A}^{2a'}=\bfZ_2$ 
$$
\t \mt -{1\over \tau} , \qquad g(\tau/2) \mt  \ 1-g(\tau/2) , \qquad 
(w_1, w_2, w_3) \mt (w_2, w_1, w_3)  \eq..
$$
Polynomial invariants: 
$$
\s_1 := w_1 + w_2, \qquad \s_2 := w_1 w_2, \qquad \S_1 := w_3  \eq..
$$
Powers and coefficient in \eliminatesingular:
$$
(P,Q,R;p,q,r;M) = (3,2,1;2,2,2; -{1\over 4})  \eq..
$$
Relation between the Halphen variables:
$$
\eqalign{
W_1 + W_3 & = \S_1 \circ (\t \mt \t/2)  \cr
W_1  - W_3  & = {2(\s_2 - \S_1 \s_1 + \S_1^2) \over \s_1 - 2\S_1}
\circ (\t \mt \t/2) \cr
 W_2 - W_3  & = {1\over 2}\s_1 -  \S_1 \circ (\t \mt \t/2)}  \eq..
$$

\noindent $ 4a' \mt 2A $: 

\nobreak
\noindent Hypergeometric identity (Goursat (138)): 
$$
F\left({1\over 8}, {1\over 8}; {3\over 4}; 4x(1-x)\right)
= F\left({1\over 4}, {1\over 4}; {3\over 4}; x\right)  \eq..
$$
Rational map: 
$$
\eqalign{
F &= 152 + G^2 \circ (\t \mt \t/2), 
\cr
f &= 4g(1-g) \circ (\t \mt \t/2)}
\eq.. 
$$
Symmetrizing group: \quad  $\bfS_{2A}^{4a'} = \bfZ_2$: 
$$
\t \mt -{1\over 2\tau}, \qquad g(\tau/2) \mt  \ 1-g(\tau/2) , \qquad 
(w_1, w_2, w_3) \mt (w_2, w_1, w_3)  \eq..
$$
Polynomial invariants: 
$$
\s_1 := w_1 + w_2, \qquad \s_2 := w_1 w_2, \qquad \S_1 := w_3  \eq..
$$
Powers and coefficient in \eliminatesingular:
$$
(P,Q,R;p,q,r;M) = (2,1,1;1,1,2;-{1\over 4})  \eq..
$$
Relation between the Halphen variables:
$$
\eqalign{
W_1 + W_3 & = \S_1\circ (\t \mt \t/2)  \cr
W_1  - W_3  & = {2(\s_2 - \S_1 \s_1 + \S_1^2) \over \s_1 - 2\S_1}
\circ (\t \mt \t/2) \cr
W_2 - W_3  & = {1\over 2}\s_1 - \S_1 \circ (\t \mt \t/2)}  \eq..
$$

\noindent $ 6a' \mt 3A $: 

\nobreak
\noindent Hypergeometric identity (Goursat (138)): 
$$
F\left({1\over 6}, {1\over 6}; {5\over 6}; 4x(1-x)\right)
= F\left({1\over 3}, {1\over 3}; {5\over 6}; x\right)  \eq..
$$
Rational map: 
$$
\eqalign{
F &= 66 + G^2 \circ (\t \mt \t/2),
\cr
f  &= 4g(1-g) \circ (\t \mt \t/2)}
\eq.. 
$$
Symmetrizing group: \quad  $\bfS_{3A}^{6a'} = \bfZ_2$: 
$$
\t \mt -{1\over 3\tau}, \qquad g(\tau/2) \mt  \ 1-g(\tau/2) , \qquad 
(w_1, w_2, w_3) \mt (w_2, w_1, w_3)  \eq..
$$
Polynomial invariants: 
$$
\s_1 := w_1 + w_2, \qquad \s_2 := w_1 w_2, \qquad \S_1 := w_3  \eq..
$$
Powers and coefficient in \eliminatesingular:
$$
(P,Q,R;p,q,r;M) = (3,1,2;1,1,4;{1\over 16})  \eq..
$$
Relation between the Halphen variables:
$$
\eqalign{
W_1 + W_3 & = \S_1 \circ (\t \mt \t/2) \cr
W_1  - W_3  & = {2(\s_2 - \S_1 \s_1 + \S_1^2) \over \s_1 - 2\S_1}
\circ (\t \mt \t/2) \cr
W_2 - W_3  & = {1\over 2}\s_1 -  \S_1 \circ (\t \mt \t/2)}  \eq..
$$

\noindent $ 4C \mt 2B $: 

\nobreak
\noindent Hypergeometric identity (Goursat (44)): 
$$
(1-z)^{-{1\over 4}}F\left({1\over 4}, {1\over 4}; 1;z:= 
{x^2\over 4(x-1)}\right) = F\left({1\over 2}, {1\over 2}; 1; x\right)  \eq..
$$
Rational map: 
$$
F = G + {2^8\over G+ 8}, \qquad f= {(g +1)^2\over 4g} 
\eq.. 
$$
Symmetrizing group: \quad $\bfS_{2B}^{4C}= \G_0(2)/ \G_0(4) =\bfZ_2$: 
$$
\t \mt {\tau \over 2\tau +1}, \qquad g \mt  \ {1\over g} , \qquad 
(w_1, w_2, w_3) \mt (w_1, w_3, w_2)  \eq..
$$
Polynomial invariants: 
$$
\s_1 := w_2 + w_3, \qquad \s_2 := w_2 w_3, \qquad \S_1 := w_1  \eq..
$$
Powers and coefficient in \eliminatesingular:
$$
(P,Q,R;p,q,r;M) = (2,1,-1;4,-1,-1;{1\over 4})  \eq..
$$
Relation between the Halphen variables:
$$
\eqalign{
W_1 + W_2  & = 2\S_1  \cr
W_1  - W_3  & =  2 \S_1 - \s_1  \cr
W_2 - W_3  & = {\s_1^2 - 4 \s_2 \over 2\S_1 - \s_1}} \eq..
$$

\noindent $ 4C' \mt 2B $: 

\nobreak
\noindent Hypergeometric identity (Goursat (138)): 
$$
F\left({1\over 4}, {1\over 4}; 1; 4x(1-x)\right)
= F\left({1\over 2}, {1\over 2}; 1; x\right)  \eq..
$$
Rational map: 
$$
\eqalign{
F &= -40 + G^2 \circ (\t \mt \t/2), \cr
f  &= (2g-1)^2 \circ (\t \mt \t/2)}
\eq.. 
$$
Symmetrizing group: \quad $\bfS_{2B}^{4C'}= \G_0(2)/ \G_(2) =\bfZ_2$: 
$$
\t \mt \tau + 1, \qquad g(\tau/2) \mt  \ 1- g(\tau/2) , \qquad 
(w_1, w_2, w_3) \mt (w_2, w_1, w_3)  \eq..
$$
Polynomial invariants: 
$$
\s_1 := w_1 + w_2, \qquad \s_2 := w_1 w_2, \qquad \S_1 := w_3  \eq..
$$
Powers and coefficient in \eliminatesingular:
$$
(P,Q,R;p,q,r;M) = (2,1,-1;2,2,-2;4)  \eq..
$$
Relation between the Halphen variables:
$$
\eqalign{
W_2 + W_3 & = \S_1 \circ (\t \mt \t/2) \cr
W_1  - W_3  & =  {1\over 2}\s_1 -  \S_1 \circ (\t \mt \t/2)\cr
W_2 - W_3  & = {2(\s_2 - \S_1 \s_1 + \S_1^2) \over \s_1 - 2\S_1}
\circ (\t \mt \t/2)}  \eq..
$$

\noindent $ 2B \mt 2A $: 

\nobreak
\noindent Hypergeometric identity (Goursat (44)):
$$
(1-z)^{-{1\over 8}}F\left({1\over 8}, {1\over 8}; {3\over 4};z:= {x^2\over
4(x-1)}\right) = F\left({1\over 4}, {1\over 4}; {1\over 2}; x\right)  \eq..
$$
Rational map: 
$$
F = G + {2^{12}\over G - 24}, \qquad f= {g^2\over  4(g-1)} 
\eq.. 
$$
Symmetrizing group: \quad $\bfS_{2A}^{2B}= \G^{(2)}_0(2)/ \G_0(2) =\bfZ_2$: 
$$
\t \mt -{1\over 2 \tau}, \qquad g \mt  \ {g\over g-1} , \qquad 
(w_1, w_2, w_3) \mt (w_3, w_2, w_1)  \eq..
$$
Polynomial invariants: 
$$
\s_1 := w_1 + w_3, \qquad \s_2 := w_1 w_3, \qquad \S_1 := w_2  \eq..
$$
Powers and coefficient in \eliminatesingular:
$$
(P,Q,R;p,q,r;M) = (2,1,1;1,2,1;4)  \eq..
$$
Relation between the Halphen variables:
$$
\eqalign{
W_1 + W_2 & = 2\S_1  \cr
W_1  - W_3  & = {\s_1^2 - 4 \s_2 \over 2\S_1 - \s_1}  \cr
W_2 - W_3  & = -\s_1 + 2 \S_1 }  \eq..
$$

\noindent $ 3B \mt 3A $: 

\nobreak
\noindent Hypergeometric identity (Goursat (44)):
$$
(1-z)^{-{1\over 6}}F\left({1\over 6}, {1\over 6}; {5\over 6};z:= {x^2\over
4(x-1)}\right) = F\left({1\over 3}, {1\over 3}; {2\over 3}; x\right)  \eq..
$$
Rational map: 
$$
F = G + {3^{6}\over G - 12}, \qquad f= {g^2\over  4(g-1)} 
\eq.. 
$$
Symmetrizing group: \quad   $\bfS_{3B}^{3A} =\bfZ_2$: 
$$
\t \mt -{1\over 3\tau}, \qquad g \mt  \ {g\over g-1} , \qquad 
(w_1, w_2, w_3) \mt (w_3, w_2, w_1)  \eq..
$$
Polynomial invariants: 
$$
\s_1 := w_1 + w_3, \qquad \s_2 := w_1 w_3, \qquad \S_1 := w_2  \eq..
$$
Powers and coefficient in \eliminatesingular:
$$
(P,Q,R;p,q,r;M) = (3,1,2;2,2,2;16)  \eq..
$$
Relation between the Halphen variables:
$$
\eqalign{
W_1 + W_2 & = 2\S_1  \cr
W_1  - W_3  & = {\s_1^2 - 4 \s_2 \over 2\S_1 - \s_1}  \cr
W_2 - W_3  & = -\s_1 + 2 \S_1 }  \eq..
$$

\noindent $ 2B \mt 1A $: 

\nobreak
\noindent Hypergeometric identity (Goursat (123)): 
$$
F\left({1\over 12}, {1\over 12}; {1\over 2};  -{x (x-9)^2\over
27(x-1)^2}\right) 
= (1-x)^{{1\over 6}}F\left({1\over 4}, {1\over 4}; {1\over 2}; x\right)  \eq..
$$
Rational map: 
$$
\eqalign{
F &= G +{2^{16}(3G+184) \over (G-24)^2}, 
\cr
f &= {(g + 3)^3\over  27(g-1)^2}} 
\eq.. 
$$
Symmetrization quotient: \quad   $\G/ \G_0(2) =\bfS_{1A}^{4C}/\bfS_{2B}^{4C}
=\bfS_3/\bfZ_2$ 
\hfill
\medskip
\noindent
Polynomial invariants: 
$$
\eqalign{
\S_1 &:= 3w_1 + 2w_2 + w_3, \qquad \S_2 := (w_1 - w_3)(4w_1-3w_2 -w_3), 
\cr
 \S_3 &:= (w_1 - w_3)^2(8w_1-9w_2 + w_3) }\eq..
$$
Powers and coefficient in \eliminatesingular:
$$
(P,Q,R;p,q,r;M) = (3,2,1;3,2,1;-27)  \eq..
$$
Relation between the Halphen variables:
$$
\eqalign{
2W_1 + 3W_2  + W_3 & =\phantom{-}\S_1  \cr
W_1  - W_3  & =   -{\S_2^2\over \S_3} \cr
W_2 - W_3  & =  -{\S_3 \over \S_2}}  \eq..
$$

\noindent $ 2B' \mt 1A $: 

\nobreak
\noindent Hypergeometric identity (Goursat (122)): 
$$
F\left({1\over 12}, {1\over 12}; {1\over 2};  {x (9-8x)^2\over
27(1-x)}\right) 
= (1-x)^{{1\over 12}}F\left({1\over 4}, {1\over 4}; {1\over 2}; x\right)  \eq..
$$
Rational map: 
$$
\eqalign{
F &= G^2 -552 +{2^{12} \over G-24} \circ (\t \mt \t/2), 
\cr
 f &= {(3-4g)^3\over  27(1-g)} \circ (\t \mt \t/2)}
\eq.. 
$$
Symmetrization quotient: \quad  $\G/ \G_0(2) =\bfS_{1A}^{4C'}/\bfS_{2B}^{4C'}
=\bfS_3/\bfZ_2$
\hfill 
\medskip
\noindent
Polynomial invariants: 
$$
\eqalign{
\S_1 &:= 3w_1 + w_2 + 2w_3, \qquad \S_2 := (w_1 - w_3)(w_1 + 3w_2 - 4w_3), 
\cr
 \S_3 &:= (w_1 - w_3)^2(w_1-9w_2 + 8w_3) }\eq..
$$
Powers and coefficient in \eliminatesingular:
$$
(P,Q,R;p,q,r;M) = (3,2,1;3,1,2;-{27\over 64})  \eq..
$$
Relation between the Halphen variables:
$$
\eqalign{
3W_1 + 2W_2  + W_3 & = {1\over 2}\S_1 \circ (\t \mt \t/2)  \cr
W_1  - W_3  & = -{\S_2^2\over 2\S_3}\circ (\t \mt \t/2)  \cr
W_2 - W_3  & =  -{\S_3 \over2 \S_2}\circ (\t \mt \t/2) }  \eq..
$$

\noindent $ 3B \mt 1A $: 

\nobreak
\noindent Hypergeometric identity (Goursat (131)): 
$$
F\left({1\over 12}, {1\over 12}; {2\over 3}; {x (x+8)^3\over
64(x-1)^3}\right) 
= (1-x)^{{1\over 4}}F\left({1\over 3}, {1\over 3}; {2\over 3}; x\right)  \eq..
$$
Rational map: 
$$
F = G+{3^9(10G^2 +732G + 9459) \over (G-12)^3} , 
\qquad f= {g(g+8)^3\over  64(g-1)^3} 
\eq.. 
$$
Symmetrization quotient: \quad   $\G/ \G_0(3) =
\bfS_{1A}^{9B}/\bfS_{3B}^{9B}=\bfA_4/\bfZ_3$\hfill
\medskip
\noindent
Polynomial invariants: 
$$
\eqalign{
\S_1 &:= 3w_1 + 2w_2 + w_3, \qquad \S_2 := (w_1 - w_3)(9w_1 - 8w_2 - w_3), 
\cr \S_3 &:=
 (w_1 - w_3)(27w_1^2 - 36 w_1 w_2 + 8w_2^2 - 18 w_1 w_3 + 2 w_2 w_3 - w_3^2)}
\eq..
$$
Powers and coefficient in \eliminatesingular:
$$
(P,Q,R;p,q,r;M) = (3,2,1;3,2,1;64)  \eq..
$$
Relation between the Halphen variables:
$$
\eqalign{
3W_1 + 2W_2  + W_3 & = \phantom{-}\S_1  \cr
W_1  - W_3  & = -{\S_2^2 \over \S_3}  \cr
W_2 - W_3  & = -{\S_3\over \S_2} }  \eq..
$$

\noindent $ 3B' \mt 1A $: 

\nobreak
\noindent Hypergeometric identity (Goursat (130)): 
$$
F\left({1\over 12}, {1\over 12}; {2\over 3}; {x (9x-8)^3\over
64(x-1)}\right) 
= (1-x)^{{1\over 4}}F\left({1\over 3}, {1\over 3}; {2\over 3}; x\right)  \eq..
$$
Rational map: 
$$
F = G^3 -162G +228 +{3^6 \over (G-12)} -744 \circ (\t \mt \t/3) , 
\quad f= {g(9g - 8)^3\over  64(g-1)} \circ (\t \mt \t/3)
\eq.. 
$$
Symmetrization quotient: \quad   $\G/ \G_0(3) =
\bfS_{1A}^{9B'}/\bfS_{3B'}^{9B'}=\bfA_4/\bfZ_3$
\medskip
\noindent
Polynomial invariants: 
$$
\eqalign{
\S_1 &:= w_1 + 2w_2 + 3w_3, \qquad \S_2 := (w_1 - w_3)(w_1 + 8w_2 - 9w_3), \cr
\S_3 &:=
 (w_1 - w_3)(w_1^2 - 20 w_1 w_2 - 8w_2^2 + 18 w_1 w_3 + 36 w_2 w_3 - 27 w_3^2)}
\eq..
$$
Powers and coefficient in \eliminatesingular:
$$
(P,Q,R;p,q,r;M) = (3,2,1;2,3,1;{64 \over 729})  \eq..
$$
Relation between the Halphen variables:
$$
\eqalign{
3W_1 + 2W_2  + W_3 & = {1\over 3}\S_1 \circ (\t \mt \t/3) \cr
W_1  - W_3  & = -{\S_2^2 \over 3\S_3}\circ (\t \mt \t/3)  \cr
W_2 - W_3  & = -{\S_3\over 3\S_2} \circ (\t \mt \t/3)}  \eq..
$$

\noindent $ 4C \mt 2a $: 

\nobreak
\noindent Hypergeometric identity:
$$
F\left({1\over 6}, {1\over 6}; {2\over 3}; {i(x+\o)^3\over
3\sqrt{3}x(1-x)}\right) 
= (x(1-x))^{{1\over 6}}F\left({1\over 2}, {1\over 2}; 1; x\right),
\qquad \o := e^{2\pi i\over 3}  \eq..
$$
Rational map: 
$$
F = G+ {2^9G\over G^2-64} , 
\qquad f= {i(g+\o)^3\over3\sqrt{3}g(1-g)}
\eq.. 
$$
Symmetrizing group: \quad  $\bfS_{2a}^{4C} =  \bfZ_3$: 
$$
\t \mt -{1\over 4\tau -2}, \qquad g \mt 1 -  {1\over g} , \qquad 
(w_1, w_2, w_3) \mt (w_3, w_1, w_2)  \eq..
$$
Polynomial invariants: 
$$
\eqalign{
\s_1 &:= w_1 + w_2 + w_3 , \cr
\s_2 &:= w_1 w_2 + w_2 w_3 + w_1 w_3, \cr
\S_3 &:= (w_1 + \o w_2 + \o^2 w_3)^3.}
\eq..
$$
Powers and coefficient in \eliminatesingular:
$$
(P,Q,R;p,q,r;M) = (1,1,1;1,1,1;3\sqrt{3} i)  \eq..
$$
Relation between the Halphen variables:
$$
\eqalign{
W_1 + W_2 + W_3 & = \s_1 \cr
W_1 - W_3  & = -{( \s_1^2 - 3\s_2)^2 \over \S_3} \cr
W_2  - W_3  & = -{\S_3 \over  \s_1^2 - 3\s_2}}  \eq..
$$

\section 3. Four vertex systems
\nobreak

\Subtitle {3a. Solutions of Generalized Halphen systems}

\nobreak
The examples listed in Table 2 below all involve Fuchsian operators 
\Fuchstand with four regular singular points (including $\infty$). We use the
following linear combination of Ohyama's dynamical variables \cite{O2} as our
phase space coordinates
$$
u:= X_0 = {1\over 2} {f'' \over f'}, \qquad v_i :={1\over 2}( X_0 - X_i) =
{1\over 2}{f'\over f- a_i} \qquad i=1,2,3,
\eq uvaidef..
$$
where $(a_1,a_2,a_3)$ are the locations of the finite poles of the rational
function $R(f)$ in the Fuchsian equation \Fuchstand and the Schwarzian 
equation \Schwarzftau.  The cases considered here involve the replicable
functions denoted: 6C, 6D, 6E, 6c, 6E, 9B in \cite{FMN}. These include all 
cases with integer $q$-series coefficients that, when composed with a
suitably  defined rational map of degree $\le 4$, give one of the triangular
functions listed in Table 1. Each provides modular solutions to a system of
the type formulated by Ohyama \cite{O2}, generalizing the equations of Halphen
type.  Further cases of such replicable functions do exist, but they may all 
be related to one of the above through an affine transformation that relocates
the finite poles of the Fuchsian operator \Fuchstand, composed with a M\"obius
transformation of the modular variable $\tau$, and hence they are not listed
separately. As in the previous section, the function $F(\tau)$ denotes the
normalized $q$--series \qseriesnorm as in \cite{FMN}, which is related by an
affine transformation \Affine to a function $f(\tau)$ normalized, if possible,
to take values  $(a, 0, 1,\infty)$ at the vertices, for some real $a<0$. In
two cases: 6D and 9B, this is not possible, because the three finite singular
points are not collinear. In case 9B, they are  normalized instead to the cube
roots of $1$, as in \cite{O3}, while in case 6D they are normalized so that
the single real pole is located at $1$. We also list, for each case,
expressions for $f(\tau)$ in terms of the $\eta$-function or null
$\vartheta$-functions and the generators $\{\r_1, \r_2,\r_3\}$ of
the automorphism group corresponding to the finite vertices of the fundamental
region normalized to have integer entries. These satisfy
$$
\r_{\infty}\r_3\r_2\r_1 = m\bfI \eq..
$$
where $\r_{\infty}$ is given by \rhoinf and $m\neq 0$ is an integer.

The differential equations satisfied by the quantities $(u, v_1, v_2,
v_3)$ are determined by a quadratic vector field together with a quadratic
constraint.  The equations for $v'_1, v'_2, v'_3$ following from \uvaidef 
are of the form
$$
v'_i= -2 v_i^2 + 2 u v_i, \qquad i=1,2,3,  \eq vaidiffeq..
$$
while the constraint equation is
$$
(a_1-a_2)v_1 v_2 + (a_2-a_3)v_2 v_3 + (a_3-a_1)v_1 v_3 = 0.
\eq vaiconstreq..
$$
When the rational function $R(f)$ appearing in eqs.~\Fuchstand and
\Schwarzftau is expressible in the form
$$
R(f) = {1\over 4} \sum_{i,j=1}^3 {r_{ij}\over (f-a_i)(f-a_j)},  \eq
Rffourvertex..
$$
the remaining equation, for $u'$, is
$$
u' = u^2 - \sum_{i,j=1}^n r_{ij} v_iv_j.  \eq ufourdiffeq..
$$

In the table, rather than listing $R(f)$ directly, we give the quadratic 
form in $(v_1, v_2, v_3)$ appearing on the RHS of eq.~\ufourdiffeq. (Note that
this quadratic form, and hence also the coefficients $r_{ij}$ appearing in
\Rffourvertex, is arbitrary up to the addition of any multiple of the 
constraint \vaiconstreq, but such a change leaves $R(f)$ invariant.) 
The general solution is again obtained by composing the function $f$ with a
M\"obius transformation \LFT. This amounts to transforming the function $2u$ 
as an affine connection
$$
u \lra {1\over (c\t + d)^2} u\circ T -{c \over c\t + d},  \eq..
$$
and the functions $v_i$ as $1$--forms
$$
v_i \lra {1\over (c\t + d)^2}v_i\circ T .  \eq..
$$
\bigskip
\bigskip
\centerline{\bf{Table 2.  Four Vertex Replicable Functions}}
\nobreak
\centerline{{\smaller (admitting a rational map of degree 
${\scriptstyle \leq 4}$ to a triangular replicable function)}}
\nobreak\medskip \smallskip 
\centerline{
\vbox{\tabskip=0pt \offinterlineskip
\def\tablerule{\noalign{\hrule}}
\halign to430pt{\strut#& \vrule#\tabskip=.5em plus1em&
 \hfil#\hfil & \vrule # &\hfil #\hfil & \vrule # &
 \hfil#\hfil & \vrule# & \hfil#\hfil & \vrule# &
 \hfil#\hfil & \vrule# & \hfil#\hfil & \vrule#
\tabskip=0pt\cr\tablerule
&& Name && $(a_1,a_2,a_3)$ && $\matrix{\rho_1 \cr \rho_2 \cr \rho_3}$ 
&& $\displaystyle{ \sum_{i,j=1}^3 r_{ij}v_i v_j} $ && $ F $ &&$f(\tau)$  
&\cr \tablerule
&& $\matrix{6C}$ && $(-3,0,1)$ && $\matrix{\smatrix[3 \& -2\\ 6 \& -3]\cr 
\smatrix[3 \& -1 \\ 12 \& -3]\cr 
\smatrix[-1 \& \phantom{-}0\\ \phantom{-}6 \& -1]}$  
&& $\matrix{{3\over 4}v_1^2 + {3\over 4} v_2^2 + v_3^2\cr
 - {1\over 2} v_2 v_3 - v_1 v_3}$ 
&& $4f + 2$  && $1 + {1\over 4} {\eta^6(\t) \eta^6(3\t) \over \eta^6(2\t)
\eta^6(6\t)}$ 
&\cr\tablerule
&& $\matrix{6D}$ && $\matrix{(\b,\bar{\b},1) \cr
\b:=-{3\over 4} + \sqrt{2}i}$ 
&&
$\matrix{ \smatrix[4 \& -3 \\ 6 \& -4]\cr
\smatrix[2 \& -1\\ 6 \& -2]\cr \smatrix[-1 \& \phantom{-}0\\
\phantom{-}6 \& -1]}$  && $\matrix{{3\over 4}v_1^2 + {3\over 4} v_2^2 + v_3^2
\cr
 + {131\over 162} v_1 v_2 \cr - {28 -16\sqrt{2}i\over 81}v_1 v_3 \cr
- {28 +16\sqrt{2}i\over 81}v_2 v_3}$ 
&& $4f$  && $1 + {1\over 4} {\eta^4(\t) \eta^4(2\t) \over \eta^4(3\t)
\eta^4(6\t)}$ 
&\cr\tablerule
&& $\matrix{6E\cr \sim \G_0(6)}$ && $(-{1\over 8},0,1)$
 && $\matrix{\smatrix[5 \& -3\\ 12 \& -7]\cr 
\smatrix[5 \& -2 \\ 18 \& -7]\cr 
\smatrix[-1 \& \phantom{-}0\\ \phantom{-} 6 \& -1]}$  
&& $\matrix{v_1^2 + v_2^2 + v_3^2\cr
 - {10\over 9} v_2 v_3 -{8\over 9} v_1 v_3}$ 
&& $8f -3$  && $1 + {1\over 8} {\eta^5(\t) \eta(3\t) \over \eta (2\t)
\eta^5(6\t)}$ 
&\cr\tablerule
&& $\matrix{6c}$ && $(-1,1,0)$ && $\matrix{\smatrix[8 \& -7 \\ 12 \& -10] \cr
\smatrix[2 \& -1\\ 12 \& -4] \cr  \smatrix[-1 \& \phantom{-}0\\ 12 \& -1]}$ &&
$\matrix{{8\over 9}v_1^2 + {8\over 9} v_2^2 + v_3^2\cr
 +{16\over 9} v_1 v_2}$ 
&& $i3\sqrt{3}f$  && $-{i\over 3\sqrt{3}} {\eta^6(2\t) \over \eta^6(6\t)}$ 
&\cr\tablerule
&& $\matrix{8E\cr \sim \G_0(8)}$ && $(-1,0,1)$ 
&& $\matrix{\smatrix[3 \& -2\\ 8 \& -5]\cr 
\smatrix[3 \& -1 \\ 16 \& -5]\cr 
\smatrix[-1 \& \phantom{-}0 \\ \phantom{-}8 \& -1]}$  
&& $\matrix{v_1^2 +  v_2^2 + v_3^2\cr - 2v_1 v_3 }$ && $4f $   
&& $\matrix{1 + {1\over 4} {\eta^4(\t) \eta^2(4\t) \over
\eta^2(2\t) \eta^4(8\t)}\cr
={\th_3^2(2\t) + \th_4^2(2\t) \over \th_3^2(2\t) - \th_4^2(2\t)}}$ 
 &\cr\tablerule
&& $\matrix{9B\cr \sim \G_0(9)}$ 
&& $\matrix{(\o,\bar{\o},1)\cr \o:= e^{2\pi i\over 3}}$ &&
$\matrix{ \smatrix[5 \& -4 \\ 9 \& -7]\cr \smatrix[2 \& -1\\ 9 \& -4]\cr 
\smatrix[-1 \& \phantom{-}0\\ \phantom{-}9 \&-1]}$  
&& $\matrix{v_1^2 +  v_2^2 + v_3^2\cr
 - v_1 v_2 \cr - (1-\o)v_1 v_3 \cr - (1-\bar{\o})v_2 v_3}$ 
&& $3f$  && $1 + {1\over 3} {\eta^3(\t) \over \eta^3(9\t)}$ 
&\cr\tablerule
\hfil\cr}}}
\medskip 
\Subtitle {3b. Rational maps to triangular cases}
\medskip
\nobreak

   In the following, we catalogue the irreducible rational maps of degree
$\le 4$ which, like those in Section 2b, relate the modular functions listed
above to the triangular cases of  Section 2a. The same notation is used as in
Section 2b, $(F,G)$ denoting a pair  of replicable functions, with $F$ given 
as a rational function of $G$,  where $F$ is one of the triangular functions of
Section 2a, with normalized $q$--series given by \qseriesnorm, and the four
vertex function $G$ is one of the cases listed in Section 3a, similarly
normalized. The corresponding pair, with values normalized to $(0,1,\infty)$ 
at the vertices in the triangular case and to the values $(a_1, a_2, a_3)$
listed in Table 2 for the four vertex case, is denoted $(f,g)$.  The
generalized Halphen variables for $f$ are again as defined in \WHalphenf,
while the corresponding ones for $g$ are defined by replacing $f$ in \uvaidef
by $g$. Taking derivatives of the rational map relating $f$ to $g$, we  obtain
relations between the associated generalized Halphen variables. Analogously to
relation
\eliminatesingular, there is always a relation of the form
$$
{f'^{P+Q+R}\over f^{P+R} (f-1)^{Q+R}} = M {g'^k \over (g-a_1)^p
(g-a_2)^q(g-a_3)^r},  \eq eliminatesingularfour..
$$
with integer powers $(P,Q,R, p,q,r,k)$ satisfying 
$$
P+Q+R = k,  \eq..
$$
and  $(P,Q,R)$  again chosen so that the LHS of \eliminatesingularfour has no
singularities at the vertices. The resulting quantity is again an analytic 
form  of weight $2k$ and, taking logarithmic derivatives of both sides of 
\eliminatesingularfour, we again obtain a linear relation between the 
generalized Halphen variables for the pair $(f,g)$
$$
P W_1 + Q W_2 + R W_3 = k u - p v_1 - q v_2 - r v_3.
\eq linearinvfour..
$$ 
Two other relations also follow from the definitions of these variables,
allowing us to determine the  quantities $(W_1,W_2,W_3)$ for the triangular
case as  simple rational symmetric functions of the variables $( u, v_1, v_2,
v_3)$ for the four vertex case, invariant under the larger automorphism group
$\grG_f$. In those cases where $\grG_g \ss \grG_f$ is a normal subgroup we
list the linear, quadratic and cubic polynomial invariants, in terms of which 
$(W_1, W_2, W_3)$ are expressed. In all cases a fourth invariant, which we do
not list separately, is provided by the vanishing quadratic form \vaiconstreq.
(A notational convention that is slightly different from that of Table 2 is
used below; the subscripts on the $v$-variable designate the location of the
poles rather than their order in the sequence $(1,2,3)$ indicated in the
second column of the table; i.e., what appears in the table  as $v_i$ is here
denoted $v_{a_i}$.)
\medskip
\noindent $ 8E \mt 4C $: 

\nobreak
\noindent Rational map: 
$$
\eqalign{
F &= G +{2^4\over G}, 
\cr
 f &= {(g + 1)^2\over 4g}} 
\eq.. 
$$
Powers and coefficient in \eliminatesingularfour:
$$
(P,Q,R;p,q,r,k;M) = (1,0,0;-1,1,1,1;1)  \eq..
$$
Symmetrizing group: \quad $\bfS_{4C}^{8E}= \G_0(8) / \G_0(4) =\bfZ_2$: 
$$
\eqalign{
\t &\mt  {-\tau\over 4\tau -1}, \qquad g \mt {1\over g} , \cr
(u, v_{-1}, v_0, v_1 ) &\mt (u-2v_0,v_1 - v_0, -v_0, v_1 - v_0)}  \eq..
$$
Polynomial invariants: 
$$
\S_1 := u + v_{-1} - v_0 -v_1, \qquad \S_1':=4v_1-2v_0 ,
\qquad \S_2 := 4v_0^2
\eq..
$$
Relation between the Halphen variables:
$$
\eqalign{
W_1 & = \S_1  \cr
W_1  - W_3  & = {\S_2 \over \S_1'} \cr
W_2 - W_3  & = \S_1' }.  \eq..
$$

\noindent $ 6c \mt 6a $: 

\nobreak
\noindent Rational map: 
$$
\eqalign{
F &= G -{3^3\over G}, 
\cr
 f &= {(g+1)^2\over 4g}} 
\eq.. 
$$
Powers and coefficient in \eliminatesingularfour:
$$
(P,Q,R;p,q,r,k;M) = (1,1,4;4,2,4,6;256)  \eq..
$$
Symmetrizing group: \quad $\bfS_{6a}^{6c} = \bfZ_2$: 
$$
\eqalign{
\t &\mt {-1\over 12 \tau -6}, \qquad g \mt {1\over g} , \cr
(u, v_{-1}, v_0, v_1 ) &\mt (u-2v_0,v_{-1} - v_0, -v_0,v_1 - v_0)}  \eq..
$$
Polynomial invariants: 
$$
\S_1 := 6u - 4v_{-1} - 2v_0  - 4v_1, \qquad \S_1' := 4v_1-2v_0, 
\qquad \S_2 := 4v_0^2
\eq..
$$
Relation between the Halphen variables:
$$
\eqalign{
W_1 & = \S_1  \cr
W_1  - W_3  & =  \S_1' \cr
W_2 - W_3  & =  {\S_2 \over \S_1'}}.  \eq..
$$
\noindent $ 6C \mt 3A $: 

\nobreak
\noindent Rational map: 
$$
\eqalign{
F &= G + {2^8(3G-2)\over (G-6)^2},
\cr
 f &= {(g+3)^3\over 27 (g-1)^2}} 
\eq.. 
$$
Powers and coefficient in \eliminatesingularfour:
$$
(P,Q,R;p,q,r,k;M) = (3,1,2;2,3,2,6;3^6)  \eq..
$$
Symmetrization quotient: \quad
$\bfS_{6C}^{12E'}/\bfS_{3A}^{12E'}=\bfS_3/\bfZ_2$ 
\medskip \noindent
Polynomial Invariants:
$$
\S_1 := 6u - 2 v_{-3} - 3 v_0 - 2 v_1, \qquad
\S_2 := v_1(4v_1 - 3 v_0), \qquad
\S_3 := v_1^2 (9v_0 -8 v_1)   \eq..
$$
Relation between the Halphen variables:
$$
\eqalign{
3W_1 +W_2 +2W_3 &  =  \S_1  \cr
W_1 - W_3 & = {2\S_2^2\over \S_3} \cr 
W_2 - W_3  & =  {2\S_3\over \S_2}.}  \eq..
$$

\noindent $ 6E \mt 3B $: 

\nobreak
\noindent Rational map: 
$$
\eqalign{
F &= G + {2^4\over (3G + 13)^2},
\cr
 f &= {(1+2g)^3\over 27 g^2}} 
\eq.. 
$$
Powers and coefficient in \eliminatesingularfour:
$$
(P,Q,R;p,q,r,k;M) = (3,1,-1;0,5,-3,3;{8\over 27})  \eq..
$$
Symmetrization quotient: \quad $\G_0(3) / \G_0(6) =
\bfS_{3B}^{12I'}/\bfS_{6E}^{12I'}=\bfS_3/\bfZ_2$ 
\medskip \noindent
Relation between the Halphen variables:
$$
\eqalign{
3W_1 + W_2 -W_3 &  =  3u - 5v_0 + 3v_1  \cr
W_1  - W_3  & =   {4(v_0 - 3 v_1)^2 \over 9v_1 - v_0} \cr
W_2 - W_3  & =  {4 v_0^2 \over 3v_1 - v_0}.}  \eq..
$$

\noindent $ 9B \mt 3B $: 

\nobreak
\noindent Rational map: 
$$
\eqalign{
F &= G + {3^3(2G + 3)\over G^2+3G + 9},
\cr
 f &= {(2+g)^3\over 9(1 + g  +g^2)}} 
\eq.. 
$$
Powers and coefficient in \eliminatesingularfour:
$$
(P,Q,R;p,q,r,k;M) = (3,1,-1;4,4,-6, 3;{1\over 9})  \eq..
$$
Symmetrizing group: \quad $\bfS_{3B}^{9B}= \G_0(3) / \G_0(9) = \bfZ_3 $ 
$$
\eqalign{
\t &\mt {\tau \over 3\tau -1}, \qquad g \mt  \ {\o g + 2\over g - \bar{\o}} ,
\cr
 (u, v_{\o}, v_{\bar{\o}}, v_1 ) &\mt 
(u-2v_{\bar{\o}}, -v_{\bar{\o}}, v_\o -v_{\bar{\o}}, v_1 - v_{\bar{\o}})}
\eq..
$$
Polynomial invariants: 
$$
\eqalign{
\S_1 &:= 3u + 6v_1 - 4v_{\o} - 4 v_{\bar{\o}}, 
\qquad \S_2 := 3(v_\o^2 - v_\o v_{\bar\o}+ v_{\bar\o}^2)
\cr \S_3 &:= ((1-\o)v_\o - (1-\bar{\o}) v_{\bar\o})^3}
\eq..
$$
Relation between the Halphen variables:
$$
\eqalign{
2W_1 +W_2 & = \, \S_1  \cr
W_1  - W_3  & = \,  {2\sqrt{3}i \S_2^2\over 3\S_3}
 \cr
W_2 - W_3  & =-{2\sqrt{3}i\S_3\over 3\S_2}. }  \eq..
$$

\noindent $ 6D \mt 2A $: 

\nobreak
\noindent Rational map: 
$$
\eqalign{
F &= G + {3^7(2G^2 + 20 G + 131)\over (G-4)^3},
\cr
 f &= {(4g+23)^4\over 4^7(g-1)^3}} 
\eq.. 
$$
Powers and coefficient in \eliminatesingularfour:
$$
(P,Q,R;p,q,r,k;M) = (2,1,1;2,2,1,4;4^7)  \eq..
$$
Symmetrization quotient: 
\quad $\bfS_{2A}^{18D'}/\bfS_{6D}^{18D'}=\bfS_4/\bfS_3$ 
\medskip \noindent
Relation between the Halphen variables:
$$
\eqalign{
W_1 +W_2 +W_3 & = 4u - 2v_{\b} - 2v_{\bar\b} - v_1 \cr
W_1  - W_3  & =  {2\sqrt{2}\left((-5i +\sqrt{2})v_\b + (5i + \sqrt{2}
)v_{\bar\b}\right)^3
\over \left((22i +\sqrt{2})v_\b + (-22i + \sqrt{2})v_{\bar\b}\right)
 \left((7i +4\sqrt{2})v_\b + (-7i + 4\sqrt{2})v_{\bar\b}\right)}
\cr 
W_2 - W_3  & = {16\sqrt{2} v_{\b} v_{\bar\b}
\left((22i +\sqrt{2})v_\b + (-22i + \sqrt{2})v_{\bar\b}\right)
\over  \left((-5i +\sqrt{2})v_\b + (5i + \sqrt{2})v_{\bar\b}\right)
\left((7i +4\sqrt{2})v_\b + (-7i + 4\sqrt{2})v_{\bar\b}\right)}.}  \eq..
$$

\noindent $ 6c \mt 2a $: 

\nobreak
\noindent Rational map: 
$$
\eqalign{
F &= G - {3^5(2G^2 + 81)\over G^3},
\cr
 f &= {(g+3)^3(g-1)\over 16g^3}} 
\eq.. 
$$
Powers and coefficient in \eliminatesingularfour:
$$
(P,Q,R;p,q,r,k;M) = (1,1,1;2,0,2,3;16)  \eq..
$$
Symmetrization quotient: \quad 
$\bfS_{6c}^{18|2'}/\bfS_{2a}^{18|2'}=\bfA_4/\bfZ_3$
\hfill\break \noindent 
(Remark: The group $18|2$ has a function field that is not
of genus $0$, and hence it does not appear in the list of replicable functions
\cite{FMN}.) 
\medskip \noindent
Relation between the Halphen variables:
$$
\eqalign{
W_1 +W_2 +W_3 & = \phantom{-}3u - 2v_{-1} - 2v_1  \cr
W_1  - W_3  & =  - {2v_0(3v_0 -4v_1)^2 \over (v_0-2v_1)(3v_0-2v_1)} \cr
W_2 - W_3  & =  -{2 (3v_0 - 2v_1)^2 \over 3v_0 - 4v_1}.}  \eq..
$$

\noindent $ 6E \mt 2B $: 

\nobreak
\noindent Rational map: 
$$
\eqalign{
F &= G + {3^3(10G^2 + 44G + 43)\over (G+4)^3},
\cr
 f &= {(8g^2 + 20g-1)^2\over (8g+1)^3}} 
\eq.. 
$$
Powers and coefficient in \eliminatesingularfour:
$$
(P,Q,R;p,q,r,k;M) = (2,1,-1;5,0,-4,2;2^{12})  \eq..
$$
Symmetrization quotient: \quad \quad $\bfS_{6E}^{18D'}/\bfS_{2B}^{18D'} $ 
\medskip \noindent
Relation between the Halphen variables:
$$
\eqalign{
W_1 +W_2 +W_3 & = \phantom{-} 2u - 5v_{-{1\over 8}} + 4v_1 \cr
W_1  - W_3  & =  - {2(v_0^2 + 18 v_0v_1 - 27 v_1^2) \over v_0 - 9v_1} \cr
W_2 - W_3  & = \, {128 v_0^2 v_1 \over (v_0 -9v_1)(v_0^2 + 18 v_0v_1 -
27 v_1^2)}.} 
\eq..
$$

\noindent $ 8E \mt 4a $: 

\nobreak
\noindent Rational map: 
$$
\eqalign{
F &= G - {2^4 (5G^2- 16)\over G(G^2-16)},
\cr
 f &= {i(g-i)^4\over 8g(g^2 -1)}} 
\eq.. 
$$
Powers and coefficient in \eliminatesingularfour:
$$
(P,Q,R;p,q,r,k;M) = (1,1,2;2,2,2,4;64)  \eq..
$$
Symmetrizing group: \quad $\bfS_{4a}^{8E}= \bfZ_4 $ 
$$
\eqalign{
\t &\mt {-1\over 8\tau -4}, \qquad g \mt {g - 1\over g + 1} \cr
(u, v_{-1}, v_0, v_1 ) &\mt (u-2v_{-1}, v_0 - v_{-1}, v_1 - v_{-1}, -v_{-1})}
\eq..
$$
Polynomial invariants: 
$$
\eqalign{
\S_1 &:= 4u - 2v_{-1} - 2v_0 - 2v_1, \qquad \S_2 := (v_0 - v_{-1} - v_1)^2 \cr 
\qquad \S_3 &:= (v_0-v_{-1}-v_1)(v_0 + i v_{-1} - i v_1)^2}
\eq..
$$
Relation between the Halphen variables:
$$
\eqalign{
W_1 +W_2 +W_3 & = \phantom{-}\S_1 \cr
W_1  - W_3  & =  -{\S_2^2\over \S_3} \cr
W_2 - W_3  & =   -{\S_3 \over \S_2}.} 
\eq..
$$

\section 4. $n+1$--Vertex Systems. Further Remarks.

\nobreak
   The systems listed in the preceding sections comprise a small subclass of 
the set of Hauptmoduls in \cite{FMN}. These functions all satisfy Schwarzian
equations of type \Schwarzftau with $R(f)$ of the form \NoverD, \Denom for
$1 \le n \le 25$ and determine solutions of generalized Halphen equations of 
the type introduced in \cite{O2}. In the case of $n$ finite poles in the
rational function $R(f)$, at points $\{a_i\}_{i=1, \dots n}$, the quantities
$\{v_i\}_{i=1, \dots n}$ defined, as in eq.~\uvaidef, by
$$
u:= X_0 = {1\over 2} {f'' \over f'}, \qquad v_i :={1\over 2}( X_0 - X_i) =
{1\over 2}{f'\over f- a_i} \qquad i=1,\dots n,
\eq uvaidefn..
$$
satisfy the set of quadratic constraints
$$
(a_i - a_j) v_i v_j + (a_j - a_k) v_j v_k
+ (a_k - a_i) v_k v_i  = 0. \qquad i,j, k = 1, \dots n,        
\eq vaiconstrn..
$$
which span an $n-2$ dimensional space of quadratic forms vanishing on the
quantities  $\{v_i\}$. They satisfy the differential equations 
$$
v'_i= -2 v_i^2 + 2 u v_i, \qquad i=1,\dots n,  \eq vaidiffeqn..
$$
while the remaining phase space variable $u$ satisfies
$$
u' = u^2 - \sum_{i,j=1}^n r_{ij} v_i v_j  \eq udiffeqn..
$$
when $R(f)$ is of the form
$$
R(f) = {1\over 4}\sum_{i,j=1}^n {r_{ij}\over (f-a_i)(f- a_j)}.  \eq
Rpartialdecomp..
$$
In \Rpartialdecomp we have the freedom of adding any linear combination of the
vanishing expressions
$$
{a_i - a_j \over (f-a_i)(f-a_j)} + {a_j - a_k \over (f-a_j)(f-a_k)} 
+{a_k - a_i \over (f-a_k)(f-a_i)}  = 0,  \eq..
$$
which amounts to adding linear combinations of the quadratic forms
of eqs.~\vaiconstrn to the RHS of \vaidiffeqn and \udiffeqn. Using this
freedom, we can always choose the decomposition in \Rpartialdecomp such that
the quadratic form in eq.~\udiffeqn defined by the coefficients $r_{ij}$ be
tridiagonal. Choosing the ordering $(a_1, \dots , a_n, \infty)$  corresponding
to a positively oriented path on the boudary of the fundamental region makes
this unique. The differential  systems are defined on the $3$-dimensional
subvariety of the space with linear coordinates $(u, v_1, \dots v_n)$ cut out
by the quadrics in \vaiconstrn, and are determined  by the set of $n+1$
quadratic forms appearing  in eqs.~\vaidiffeqn, \udiffeqn, defined modulo
those in eq.~\vaiconstrn. We note that the $n$ quadratic forms appearing in
eq.~\vaidiffeqn are independent of the values of the parameters, while those
defining the constraints \vaiconstrn depend only on the locations
$\{a_i\}_{i=1\dots n}$ of the poles. The diagonal coefficients of the
quadratic form in \udiffeqn are related to the angles
$\{\a_i \pi\}_{i=1 \dots n}$ at the $n$ finite vertices of the fundamental
polygon by
$$
r_{ii} = 1 -\a_i^2 .  \eq..
$$
In the case of real poles the off diagonal coefficients are determined, modulo
the vanishing quadratic forms \vaiconstrn, by the values of the {\it accessory
parameters} \cite{GS} of the corresponding mapping of the fundamental polygon
to the upper half plane.

   As an illustration of a system  associated to one of the higher level 
functions, consider the case 72e which, taken in the normalization of
\cite{FMN}, may be expressed as a ratio of $\eta$-functions
$$
f = {\eta(24\tau) \eta (36 \tau) \over \eta(12\tau) \eta(72\tau)}.  \eq..
$$
The fundamental region has $25$ finite vertices, the largest number amongst
the replicable functions. These are mapped in the $f$--plane to the origin
$f=0$ and to the twelfth roots of unity times the two reciprocal 
radii $(\sqrt{2}+1)^{1\over 3}$ and $(\sqrt{2}-1)^{1\over 3}$. The rational
function $R(f)$ entering in eqs.~\Fuchstand and \Schwarzftau is given by
$$
R(f) = {1\over 4f^2} +
{864f^{10}(f^4+1)^2(f^8-f^4+1)^2\over
(f^6+2f^3-1)^2(f^6-2f^3-1)^2(f^{12}+6f^6 +1)^2}.
\eq RFseventytwoe..
$$
 We denote the finite poles as
$$
\eqalign{
a_0 := 0, \quad
a_m := e^{(m-1)\pi i\over 6}(\sqrt{2}+1)^{1\over 3}, \quad
a_{12+m} &:= e^{(m-1)\pi i\over 6}(\sqrt{2}-1)^{1\over 6}, \qquad m=1, \dots
12.}  
\eq..
$$
 In the notation defined above, the $25$ functions $\{v_0, v_m\}_{m=1 \dots
24}$ satisfy the usual constraint equations \vaiconstrn, and equation
\vaidiffeqn for the derivatives of the $v$--variables. The quadratic form
entering in the eq.~\udiffeqn for $u'$ is
$$
\sum_{i,j=1}^nr_{ij}v_i v_j = v_0^2 + {3\over 4} \sum_{m=1}^{24} v_m^2 
- {3\over 8} \sum_{m=1}^{11}(1-e^{m\pi i\over 6})
\left((2 +\sqrt{2} )v_m v_{m+1} + (2 -\sqrt{2} )v_{12+m} v_{13+m} \right). 
\eq..
$$
The two angles in the fundamental polygon at the vertices mapping to $0$ and
$\infty$ therefore vanish, while the others are $\pi/2$. The simplicity of 
this expression is due to the invariance of the Schwarzian derivative under 
the transformation $\tau\mt \tau +{1\over 12}$, which generates the cyclic
group action 
$$
f\mt e^{\pi i\over 6} f, \qquad R(f) \mt e^{-{\pi i\over 3}} R(f),  \eq..
$$
and to the inversion symmetry
$$
R({1/f}) = f^4 R(f).   \eq..
$$

   As a final remark we note that, for the general $n$ finite pole case, when
$R(f)$ is of the form \Rpartialdecomp there is an equivalent way of expressing 
the Schwarzian equation \Schwarzftau and the associated system
\vaiconstrn--\udiffeqn in terms of an unconstrained dynamical system on the
$SL(2,\bfC)$ group manifold. To do this, let
$$
g(\tau) := \pmatrix { A & B \cr C & D} \in SL(2,\bfC), \quad 
AD-BC  =1  \eq groupel..
$$
denote an integral curve in $SL(2,\bfC)$ for the equation
$$
g' = \pmatrix{ 0 & \g \cr -1  & 0 } g,  \eq..
$$
where
$$
\g := -{1\over 4C^2} \sum_{i,j =1}^n {b_{ij}\over (Ca_i + D)(C a_j + D)}
= -{1\over C^2} R\left(-{D \over C}\right). 
\eq..
$$
Defining $f(\tau)$ to be
$$
f := - {D \over C},  \eq..
$$
it follows that this satisfies \Schwarzftau, and that
$$
u= {A\over C}, \qquad v_{a_i} = {1\over 2C(Ca_i +D)}, \qquad i=1, \dots n
\eq..
$$
satisfy the system \vaiconstrn--\udiffeqn . Equivalently, Ohyama's variables
\Ohyamavars are obtained by applying $g$ as a linear fractional transformation
to $\{\infty, a_1, \dots a_m\}$
$$
X_0 = {A\over C}, \qquad X_i = {A a_i + B \over C a_i + D},
 \qquad i=1, \dots n.  \eq..
$$ 
Conversely, up to a choice of branch in $(f')^{1\over 2}$ (which does not
affect the projective class of $g$), we may always express a solution $f$ of
\Schwarzftau in this way by defining $g$ to be
$$
g = {i \over 2(f')^{3\over 2}}\pmatrix{ f'' & 2f'^2 - f f'' \cr
                                        2f'  & - 2f f'}.  \eq..
$$

  A number of further questions suggest themselves in relation to this work.
The first concerns the origin of the Fuchsian equation associated to each of
these Hauptmoduls. For the case $\lambda(\tau)$ discussed in the introduction,
the elliptic integral formulae \ellipticinta, \ellipticintb show that the
associated hypergeometric equations are Picard--Fuchs equations
corresponding to the family of elliptic curves parametrized, in Jacobi's form,
by $\lambda(\tau)$. A similar interpretation was found by Ohyama \cite{O3} for
the $4$--vertex example denoted here as $9B$, for which the underlying Fuchsian
equation is a Picard--Fuchs equation for the Hesse pencil of elliptic curves.
All the hypergeometric functions have Euler integral representations, but it 
is not clear whether their equations might similarly be interpreted in terms 
of parametric families of elliptic curves. The same question may be asked for 
the other Hauptmoduls considered here. Another question that naturally arises
is: what are the analogues of the Chazy equation that are satisfied by the
logarithmic derivatives of the analytic forms entering in
eqs.~\eliminatesingular and \eliminatesingularfour, and how may these forms 
be expressed in terms of some standard set such as, e.g., the modular
discriminant appearing in eq.~\dlogDelta, Eisenstein series, or simply as
ratios of
$\vartheta$--functions? 
\bigskip\bigskip  
\noindent{\it Acknowledgements.}
The authors would like to thank S.~Norton, Y.~Ohyama and A.~Sebbar
for helpful discussions.   This research was supported in part by the Natural
Sciences and Engineering Research Council of Canada and the Fonds FCAR du
Qu\'ebec. 

\bigskip \bigskip
\goodbreak

\noindent
{\sectionfont A. Appendix}
\smallskip
\nobreak
We recall here the standard definitions \cite{A, F, WW} of the functions
$\vartheta_2, \vartheta_3, \vartheta_4$ and $\eta$, and a  number of
properties and relations between them that are helpful in verifying some of
the formulae of Sections 2 and 3. 
\medskip
\noindent
Dedekind $\eta$-function:
$$
\eta(\tau) := q^{1\over 24} \prod_{n=1}^\infty (1- q^n), 
\qquad (q:= e^{2\pi i\tau})  \eqno(A.1)
$$
Null $\vartheta$--functions:
$$
\eqalignno{\vartheta_2 (\tau) 
&:= \sum_{n=-\infty}^\infty q^{{1\over2}(n+{1\over 2})^2} 
= 2 q^{1\over 8} \prod_{n=1}^\infty (1 - q^n)(1 + q^n)^2  &(A.2) \cr
\vartheta_3 (\tau) &:= \sum_{n=-\infty}^\infty q^{{1\over2} n^2}
= \prod_{n=1}^\infty (1 - q^n) (1 + q^{n - {1\over 2}})^2
&(A.3) \cr
\vartheta_4 (\tau) &:= \sum_{n=-\infty}^\infty (-1)^n q^{{1\over2} n^2}
= \prod_{n=1}^\infty (1 - q^n) (1 - q^{n - {1\over 2}})^2. &(A.4) }
$$
Relations between $\vartheta_2$, $\vartheta_3$, $\vartheta_4$, and $\eta$: 
$$
\eqalignno{
\vartheta_2(\tau) & = 2 {\eta^2(2\tau) \over \eta (\tau)}    &(A.5)
\cr
\vartheta_3(\tau) & = {\eta^5(\tau) \over \eta^2(2\tau)\eta^2(\tau/2)} &(A.6)
\cr
\vartheta_4(\tau) & =  {\eta^2(\tau/2) \over \eta (\tau)}   &(A.7) \cr
\eta^3(\tau) & = {1\over 2}\vartheta_2(\tau) \vartheta_3(\tau)
\vartheta_4(\tau) &(A.8) }
$$
Modular transformations of $\vartheta_2$, $\vartheta_3$, $\vartheta_4$, and 
$\eta$: 
$$
\eqalignno{
\vartheta_2(\tau +1) & = e^{i\pi\over 4}\vartheta_2(\tau),  
\qquad \vartheta_2(-1/ \tau) = (-i\tau)^{1\over 2} \vartheta_4(\tau)   &(A.9)
\cr
\vartheta_3(\tau +1) & = \vartheta_4(\tau),  
\qquad \vartheta_3(-1/ \tau) = (-i\tau)^{1\over 2} \vartheta_3(\tau)  
&(A.10)
\cr
\vartheta_4(\tau +1) & = \vartheta_3(\tau),  
\qquad \vartheta_4(-1/ \tau) = (-i\tau)^{1\over 2} \vartheta_{2}(\tau)  
&(A.11)
\cr
\eta(\tau +1) & = e^{i\pi\over 12}\eta(\tau),  
\qquad \eta(-1/\tau) = (-i\tau)^{1\over 2} \eta(\tau)   &(A.12)
 }
$$
Identities satisfied by $\vartheta_2$, $\vartheta_3$, $\vartheta_4$, $\eta$:
$$
\eqalignno{
 \vartheta_3^4(\tau) & = \vartheta_2^4(\tau) + \vartheta_4^4(\tau)  &(A.13) 
\cr
2\vartheta_3^2(2\tau) & = \vartheta_3^2(\tau) + \vartheta_4^2(\tau)  &(A.14) 
\cr
\vartheta_2^2(\tau) & = 2\vartheta_2(2\tau)\vartheta_3(2\tau)  &(A.15) \cr
\vartheta_4^2(2\tau) & = \vartheta_3(\tau)\vartheta_4(\tau)  &(A.16)  \cr
\eta(\tau + \ \scriptstyle{{1\over 2}}) & =  {e^{\pi i\over 24} \eta^3(2\tau)
\over\eta(\tau)\eta(4\tau) } &(A.17)  \cr
\eta^3(\tau +{\scriptstyle{1\over 3}}) &=
 e^{\pi i \over 12}\eta^3(\tau) - 3\sqrt{3} e^{-{\pi i\over 12}}\eta^3(9\tau) 
&(A.18)}
$$
Differential relations satisfied by $\vartheta_2$, $\vartheta_3$, 
$\vartheta_4$:
$$
\eqalignno{
{\vartheta_2' \over \vartheta_2} - {\vartheta_3' \over \vartheta_3} & =
{i \pi \over 4} \vartheta_4^4  &(A.19) \cr
{\vartheta_3' \over \vartheta_3} - {\vartheta_4' \over \vartheta_4} & =
{i \pi \over 4} \vartheta_2^4 &(A.20) \cr
{\vartheta_2' \over \vartheta_2}- {\vartheta_4' \over \vartheta_4} & =
{i \pi \over 4} \vartheta_3^4. &(A.21)}
$$

\goodbreak
\bigskip \bigskip
\references

A& Apostol, Tom M., ``Modular Functions and Dirichlet Series in Number
Theory'', 2nd ed.   (Graduate Texts in mathematics 41, Springer-Verlag, New
York, 1976, 1990).

AH& Atiyah, M.F., and Hitchin, N.J., {\it The Geometry and Dynamics of 
Magnetic Monopoles}, Princeton University Press, Princeton (1988).

B& Brioschi, M., , ``Sur un syst\`eme d'\'equations diff\'erentielles'',
{\it C.~R.~Acad.~Sci.~Paris\/} {\bf 92}, 1389--1393 (1881).

C1& Chazy, J., ``Sur les \'equations diff\'erentielles dont l'int\'egrale
g\'en\'erale poss\`ede une coupure essentielle mobile'', {\it
C.~R.~Acad.~Sc.~Paris}, {\bf 150}, 456--458 (1910). 

C2& Chazy, J., ``Sur les \'equations diff\'erentielles du troisi\`eme ordre et
ordre sup\'erieur dont l'int\'egrale g\'en\'erale a ses points crtiques 
fixes'', {\it Acta Math.\/} {\bf 34}, 317--385 (1911). 

CAC& Chakravarty, S., Ablowitz, M.J., and Clarkson, P.A., ``Reductions of
Self-Dual Yang-Mills Fields and Classical Systems'', {\it Phys. Rev. Let.} 
{\bf 65}, 1085--1087 (1990). 

CN& Conway, J. and Norton, S.~P., ``Monstrous moonshine'', {\it
Bull.~Lond.~Math.~Soc.} {\bf 11}, 308--339 (1979).

De& Dedekind, ``\"Uber die elliptischen Modul--Functionen'', {\it J. Reine
Angew. Math.} {\bf 83}, 34--292 (1877).

Du& Dubrovin, B.A., ``Geometry of $2D$ topological field theories'', Lecture
Notes in Math. {\bf 1620}, Springer-Verlag, Berlin, Heidelberg,  New York
(1996). 

Da& Darboux, G. {\it Le\c cons sur les syst\`emes othogonaux} (2nd ed.).
Gauthiers-Villars, Paris (1910).

Fa&  Faber, ``\"Uber polynomische Entwickelungen'', {\it Math. Annalen} {\bf
57}, 389--408  (1903).

Fo& Ford, L.,  {\it Automorphic functions}, Chelsea, New York (1951).

FMN& Ford, D., McKay, J.,  and Norton, S., ``More on replicable functions''
{\it Comm.~in Algebra} {\bf 22}, 5175--5193 (1994).

GS& Gerretson, J.,  Sansone, G., {\it Lectures on the Theory of Functions of a
Complex Variable. II. Geometric Theory}. Walters--Noordhoff, Gr\"oningen (1969).

Go& Goursat, E. ``Sur L'\'Equation diff\'erentielle lin\'eaire qui admet pour
int\'egrale la s\'erie hyperg\'eom\'etrique'', {\it Ann.~Sci.~de l'\'Ecole
Normale Sup\'erieure}, {\bf X} suppl., 1--142 (1881).

GP& Gibbons, G.W., and Pope, C.N., ``The Positive Action Conjecture and
Asymptotically Euclidean Metrics in Quantum Gravity'',  {\it
Commun.~Math.~Phys.} {\bf 66}, 267--290 (1979). 

Ha& Halphen, G.-H., ``Sur des fonctions qui proviennent de l'\'equation de
Gauss'',  {\it C. R. Acad. Sci. Paris} {\bf 92}, 856--858 (1881);
 ``Sur un syst\`eme d'\'equations  diff\'erentielles'',
{\it ibid.} {\bf 92}, 1101--1103 (1881);
``Sur certains syst\`emes d'\'equations diff\'erentielles'',
{\it ibid.} {\bf 92}, 1404--1406 (1881).

H1&  Hille, Einar
{\it Ordinary Differential equations in the Complex Domain} ,  (Dover, New
York  1976), Sec. 7.3, Ch.~10. 

H2&  Hille, Einar
{\it Analytic Function Theory},  (Chelsea, New York  1973)

J& Jacobi, C.G.J., ``\"Uber die Differentialgleichung, welcher die Reihen 
$1\pm 2q + 2q^4 \pm {\rm etc.},  2^4 \sqrt{q} + 2^4 \sqrt{q^9} +
 2^4 \sqrt{q^{25}} + {\rm etc.}$ Gen\"uge Leiste'', {\it J. Reine Angew.
Math.} {\bf 36}, 97--112 (1848). (Ges. Math. Werke Bd. 2 171--191).

O1& Ohyama, Yousuke, ``Differential Relations of theta Functions'', {\it Osaka
J.~Math.} {\bf 32}, 431--450 (1995).

O2& Ohyama, Yousuke, ``Systems of nonlinear differential equations related to
second order linear equations'', {\it Osaka J.~Math.} {\bf 33},
927--949 (1996).

O3& Ohyama, Yousuke, ``Differential equations for modular forms with level
three'', Osaka Univ. ~preprint (1997).

T& Takhtajan, Leon, A. ``Modular Forms as $\tau$--Functions for 
Certain Integrable Reductions of the Yang-Mills Equations'', in: {\it
Integrable Systems. The Verdier Memorial Conference} (ed. O. Babelon, P.
Cartier and Y. Kosmann-Schwarzbach),  {\it Progress in 
Mathematics} {\bf 115}, 115--130, Birkh\"auser, Boston (1993). 

Ta& Takeuchi, K., ``Arithmetic triangle groups'', {\it J.~Math.~Soc.~Japan} 
{\bf 29}, 91--106 (1977).

WW& Whittaker, E.T., and Watson, G.N., {\it A Course in Modern Analysis},
Chapt.~21,  Cambridge University Press, 4th ed., London, N.Y. (1969).

\endreferences
\vfil

\end